\documentclass[12pt,preprint]{aastex}
\slugcomment{To appear in The Astrophysical Journal}

\lefthead{Skinner et al.}
\righthead{NGC 2071}

\begin{document}

\title{Chandra and Spitzer Imaging of the Infrared Cluster \\
        in NGC 2071}

\author{Stephen L. Skinner and Kimberly R. Sokal}
\affil{CASA, Univ. of Colorado, Boulder, CO, USA 80309-0389 }

\author{S. Thomas  Megeath}
\affil{Ritter Observatory, Dept. of Physics and Astronomy,
       University of Toledo, Toledo, OH 43606}

\author{Manuel  G\"{u}del}
\affil{Institute of Astronomy, ETH Z\"{u}rich,
   Wolfgang-Pauli-Str. 27,
   8093 Z\"{u}rich,
   Switzerland}

\author{Marc Audard}
\affil{Integral Science Data Centre, Ch. d'Ecogia 16, CH-1290 Versoix,
Switzerland}
\affil{Geneva Observatory, University of Geneva, Ch. des Maillettes 51,
       1290 Sauverny, Switzerland}

\author{Kevin M. Flaherty and Michael R. Meyer}
\affil{Steward Observatory, The University of Arizona, 
        933 N. Cherry Ave., Tucson,
	AZ, USA 85721-0065}


\author{Augusto Damineli}
\affil{Instituto de Astronomia, Geof\'{i}sica e Ci\^{e}ncias
        Atmosf\'{e}ricas,
        Universidade de S\~{a}o Paulo, Rua do Mat\~{a}o 1226, Cidade
        Universit\'{a}ria, 05508-900 S\~{a}o Paulo, SP, Brazil}

%
\newcommand{\ltsimeq}{\raisebox{-0.6ex}{$\,\stackrel{\raisebox{-.2ex}%
{$\textstyle<$}}{\sim}\,$}}
%
\newcommand{\gtsimeq}{\raisebox{-0.6ex}{$\,\stackrel{\raisebox{-.2ex}%
{$\textstyle>$}}{\sim}\,$}}

\begin{abstract}
We present results of a sensitive {\em Chandra}  X-ray observation
and {\em Spitzer} mid-IR observations of  the infrared cluster 
lying north of the NGC 2071 reflection nebula
in the Orion B molecular cloud. We focus on the dense
cluster core known as  NGC 2071-IR which   contains at 
least nine IR sources within a  40$''$ $\times$ 40$''$  region. 
This region shows clear  signs of active star formation including
powerful molecular outflows, Herbig-Haro objects,  and both
OH and H$_{2}$O  masers. We use {\em Spitzer} IRAC images
to aid in X-ray source identification and to determine
YSO classes using mid-IR colors. {\em Spitzer} IRAC colors
show that the luminous source IRS 1 is a class I  protostar.
IRS 1 is  believed  to be driving a  powerful bipolar 
molecular outflow and may be an embedded B-type star
or its progenitor.  Its X-ray spectrum
reveals a fluorescent Fe emission line at 6.4 keV, 
arising in cold material near the protostar. The line is present even in 
the absence of large flares, raising questions about the nature of
the ionizing mechanism responsible for producing the 6.4 keV
fluorescent line.
{\em Chandra} also detects X-ray sources at or near the positions of
IRS 2, IRS 3, IRS 4, and IRS 6 and a  variable X-ray source  coincident 
with the  radio source VLA 1, located just 2$''$ north of IRS 1. 
No IR data are yet available to determine a YSO classification for VLA 1,
but its high X-ray  absorption shows that it is even more deeply-embedded
than IRS 1, suggesting that it could be an even younger, less-evolved
protostar.  
\end{abstract}



%


\section{Introduction}
The Lynds 1630 dark cloud (Lynds 1962) is located 
approximately 4$'$ north of the optical  reflection nebula 
NGC 2071 in the Orion B molecular cloud (Maddalena et al. 1986).
L1630 was recognized
as a site of recent star formation by Strom et al. (1975),
who identified a young stellar population with an 
estimated age of a few million years containing 
early B and A-type stars, optically-revealed H$\alpha$
emission line stars (T Tauri stars), and several embedded 
infrared sources. An overview of star formation in
the NGC 2071 region can be found in Gibb (2008).

 The L1630 dark cloud was mapped at
2.2 $\mu$m by Strom, Strom, \& Vrba (1976) and 70
sources were detected  down to a limiting magnitude 
K $\approx$ 10.7 mag. About one-third of these lacked
optical counterparts and they estimated an extinction
A$_{v}$ $\approx$ 20 mag toward the cloud center.
A more sensitive 2.2 $\mu$m survey of L1630 was undertaken by 
Lada et al. (1991) who identified 912 sources down to 
K $\approx$ 13.0 mag, about half of which were associated
with L1630 itself.

Using known MK spectral types and optical photometry for 
a large sample of non-variable  B star members, 
the distance to this region
was estimated to be  390 pc by Anthony-Twarog (1982).
But,   the near and far edges of the Orion OB1 association 
span a range of $\approx$320 pc to 500 pc 
(Brown, de Geus, \& de Zeeuw 1994; Wilson et al. 2005). 
Using {\em Spitzer}
IRAC and MIPS data along with low-resolution optical spectra,
Flaherty \& Muzerolle (2008) determined a cluster age of
2 $\pm$ 1.5 Myr and concluded that 79\% of the late-type cluster
members in their sample have infrared excesses. They found 
evidence for active accretion in all of the IR-excess objects.

The extended NGC 2071 cluster identified by Lada et al. (1991)
covers an area of $\approx$80 arcmin$^{2}$ and contains  a 
dense subgroup  known as NGC 2071-IR (Figs. 1 and 2) which
harbors at least nine IR sources
in a 40$''$ $\times$ 40$''$ core region 
(Walther, Geballe, \& Robson 1991; Walther et al. 1993, hereafter 
W93; Tamura et al. 2007, hereafter T07). 
The position of NGC 2071-IR within NGC 2071 is shown in
Figure 2 of W93 and a wide-field perspective showing
the location of this region within the Orion complex
can be found in Fig. 3 of Maddalena et al. (1986).
Within NGC 2071-IR, the bright infrared source IRS 1
is of particular interest.  The
nature of this object is unclear but previous studies
have concluded it may be a young embedded B-type star
(W93; Snell \& Bally 1986). The NGC 2071-IR subgroup shows clear 
signs of active star-formation including powerful bipolar 
molecular outflows (Bally 1982; Snell et al. 1984; 
Aspin, Sandell, \& Walther 1992; Eisl\"{o}ffel 2000),
Herbig-Haro objects (Zhao et al. 1999),  and water masers
near IRS 1 and IRS 3 (Torrelles et al. 1998) and 
IRS 2 (Sandell et al. 1985). An OH maser 
has been detected within $\approx$20$''$ of  IRS 1 
(Johansson et al. 1974; Pankonin et al. 1977; Sandell et al. 1985) 
but the maser position is not well-enough known to determine if
it is coincident with IRS 1. {\em VLA} maps in NH$_{3}$
lines reveal what appears to be a large ring structure or edge-on
disk surrounding NGC 2071-IR (Zhou, Evans, \& Mundy 1990).

More than one outflow may  be present in NGC 2071-IR.
Using H$_{2}$ 1-0 S(1) images, 
Eisl\"{o}ffel (2000) identified at least three outflows.
These include  an east-west outflow that  may be driven by
IRS 1 (see also Aspin, Sandell, \& Walther 1992), and 
further support for such an outflow comes from high-velocity  
knots  revealed in recent 3.6 cm  {\em VLA} images 
(Carrasco-Gonz\'{a}lez et al. 2007).
Eisl\"{o}ffel (2000) also concluded that there is 
an outflow oriented in the 
northeast-southwest direction that may be driven by
IRS 3,  and a third outflow  close to IRS 7. Additional 
evidence  that IRS 3 is driving an outflow comes from
extended jet-like structure visible in high-resolution
{\em VLA} radio continuum images (Torrelles et al. 1998). 
Aspin et al. (1992) 
argued that IRS 2 may also be driving a compact bipolar
outflow. More recently, Stojimirovi\'{c} et al. (2008)
have reported evidence for at least three outflows in
the NGC 2071 region based on moderate resolution
(HPBW $\approx$45$''$) $^{12}$CO and $^{13}$CO $J$ = 1 $\rightarrow$ 0
molecular line maps. 
However, the association of individual outflows
with specific driving sources remains somewhat uncertain due
to the complex morphology of the region.

A previous X-ray observation of the NGC 2071 region with 
{\em XMM-Newton}  detected a prominent X-ray source
at a position within 1$''$ of IRS 1 
(Skinner et al. 2007, hereafter S07).
The X-ray source was identified with IRS 1, but 
was subject to limitations on source identification 
in the crowded cluster core  imposed 
by  {\em XMM}'s $\approx$4$''$ angular resolution. 
We present here the results of a higher angular 
resolution {\em Chandra} X-ray observation of NGC 2071-IR.
Our main objectives were to confirm the previous X-ray detection 
of IRS 1 with improved positional accuracy and identify
possible mechanisms for its X-ray emission. Although
the X-ray emission of optically-revealed OB stars has
been extensively studied and is thought arise in shocked winds,
much less is known about X-ray production in very young embedded 
high-to-intermediate mass stars such as IRS 1.
We confirm that
IRS 1 is indeed an X-ray source and discuss its
X-ray properties along with several other
X-ray detections in the NGC 2071-IR subgroup (IRS 2, IRS 3, 
IRS 4, and IRS 6). {\em Chandra}'s sharp angular resolution
also reveals a new X-ray source  associated
with the radio continuum source VLA 1 lying 2.$''$2 north
of IRS 1.  The VLA 1 X-ray source is  seen through 
very high absorption equivalent to A$_{\rm  V}$ $\sim$
40 mag, suggesting that it is a young heavily-embedded 
protostar. We also use {\em Spitzer} observations
and previous near-IR observations to identify IR
counterparts to the X-ray sources and  determine
their IR spectral energy distributions. We also use IR 
color-color diagrams to assign young stellar object (YSO) 
classes.

\section{Observations}

\subsection{Chandra Observations}

The {\em Chandra} observation (ObsID 7417) began on
2007 November 6 at 19:40:19 TT and ended on November 7 at 15:18:03 TT.
Pointing was centered near NGC 2071 IRS 1 at nominal pointing
coordinates (J2000) RA $=$ 05$h$ 47$m$ 04.94$s$, 
Dec. $=$ $+$00$^{\circ}$ 22$'$ 11.8$''$. 
The exposure live time was 67,180 s. 
Exposures were obtained using the ACIS-I (Advanced CCD 
Imaging Spectrometer) imaging array in faint 
timed-event mode with  3.2 s frame times. 
ACIS-I has a combined field of view (FOV) of $\approx$16.9$'$ x 16.9$'$ 
consisting of four front-illuminated 1024 x 1024 pixel CCDs with a pixel size 
of 0.492$''$. Approximately 90\% of the encircled energy at 1.49 keV
lies within 2$''$ of the center pixel of an on-axis  point source. 
Our discussion here  focuses on ACIS-I, 
but the S2 and S3 CCDs in the ACIS-S array were also enabled. 
More information on {\em 
Chandra} and its instrumentation can be found in the {\em Chandra} Proposer's 
Observatory Guide (POG)\footnote {See http://asc.harvard.edu/proposer/POG}.

The Level 1 events file provided by the {\em Chandra} X-ray
Center (CXC) was  processed using CIAO 
version 3.4\footnote{Further information on 
{\em Chandra} Interactive
Analysis of Observations (CIAO) software can be found at
http://asc.harvard.edu/ciao.} using standard science 
threads. The  CIAO processing applied calibration updates
(CALDB vers. 2.26), selected good event patterns, 
and determined source centroid positions.

Additionally, we used the CIAO {\em wavdetect} tool to
identify X-ray sources on the ACIS-I array. 
We ran {\em wavdetect} on full-resolution 
images using events in the 0.3-7 keV
range to reduce the background. The {\em wavdetect}
threshold was set at $sigthresh$ = 1.5 $\times$ 10$^{-5}$ 
and scale sizes of 1, 2, 4, 8, and 16 were used.
We identified 207 X-ray sources on ACIS-I down
to a threshold of 5 counts, including the 
sources of primary interest in this work located in the 
NGC 2071-IR core region listed in Table 1.

The unabsorbed X-ray luminosity 
upper limit for non-detections is log L$_{\rm X}$ (0.3 - 7 keV) 
$\leq$ 28.7 ergs s$^{-1}$. This upper limit from the 
Portable Interactive Multi-Mission Simulator 
(PIMMS)\footnote{Information on PIMMS can be found at
http://asc.harvard.edu/ciao/ahelp/pimms.html~.}
assumes a 5 count
on-axis detection threshold and an absorbed isothermal optically thin
plasma spectrum  similar to that observed for IRS 1 
(N$_{\rm H}$ = 2 $\times$ 10$^{22}$ cm$^{-2}$, kT = 2 keV).
At this upper limit, the observation was sensitive enough
to detect low-mass T Tauri stars in Orion down to 
$\approx$0.1 - 0.2 M$_{\odot}$, based on the known correlation
between L$_{\rm X}$ and stellar mass in T Tauri stars
(Preibisch et al. 2005). The above upper limit should be
considered as a representative value since the detection
limit depends on many factors including the spectral properties
of the source and extent to which the source is displaced off-axis.

At the sensitivity of our observation, we expect 
$\approx$1 extragalactic background source in a region 
spanning $\approx$1.6 arcmin$^{2}$, 
or $<$1 extragalactic source in the $\approx$40$''$ $\times$
40$''$ NGC 2071-IR region analyzed here. This estimate is
based on hard-band (2 - 8 keV) number counts from 
the {\em Chandra} deep field (CDF) and other X-ray surveys
(e.g. Brandt et al. 2001; Moretti et al. 2003)
and an assumed power-law spectrum with a photon power-law
index $\Gamma$ = 1.4 for extragalactic sources.
The above estimate should be considered as approximate
because the accuracy of the log $N$ - log $S$ distribution
for extragalactic sources based on deep-field observations 
when applied to lower galactic latitude star-froming regions is
not well-known. Getman et al. (2005) concluded that
CDF number counts overestimated
the number of extragalactic sources in the deep {\em Chandra}
COUP observation of the Orion Nebula Cluster by a 
factor of $\sim$3 - 4.

CIAO {\em psextract} was used to extract  source and background spectra 
for brighter sources, along with source-specific 
response matrix files (RMFs) and auxiliary
response files (ARFs) used in spectral fitting.
Light curves of brighter sources were also 
extracted using CIAO tools. We used the 3$\sigma$ source ellipses from
{\em wavdetect} to define the extraction regions and background was
extracted from adjacent source-free regions. Background is negligible,
contributing only 0.21 counts per pixel (0.3 - 7 keV) for the total
exposure time, or 
less than one count in the regions used to extract source spectra
and light curves. Spectral fitting and timing analysis were undertaken
with the HEASOFT 
{\em Xanadu}\footnote{http://heasarc.gsfc.nasa.gov/docs/xanadu/xanadu.html.}
software package including XSPEC vers. 12.4.0.
We also applied  the Kolmogorov-Smirnov (KS) test (Press et al. 1992) 
to unbinned photon arrival times to check for X-ray variability.

Since conventional spectral fitting is only practical for brighter
sources, we used the quantile analysis approach of Hong et al. (2004)
to quantify source hardness. This technique has the advantage that
it can be applied to sources that are too faint for spectral fitting. 
The quantile method uses sorted photon event energy
lists to compute the median energy E$_{50\%}$ and the quartile energies
E$_{25\%}$ and  E$_{75\%}$. The quartile energies are defined such that
one-fourth  of the counts have energies below  E$_{25\%}$, and
three-fourths below E$_{75\%}$. The quantile fractions Q$_{x}$ are computed
from quartile energies using 
Q$_{x}$ = (E$_{x\%}$ - E$_{lo}$)/(E$_{up}$ - E$_{lo}$),
where E$_{lo}$ and E$_{up}$ are the lower and upper energy bounds
of the extracted events (0.3 and 7.0 keV in this work).
Quantile values (or functions thereof) are then overplotted on
a  grid of neutral hydrogen absorption column density (N$_{\rm H}$)
and plasma energy  (kT) values generated from spectral 
simulations based on a one-temperature (1T) APEC optically thin 
thermal plasma model. This provides a graphical representation of the 
source hardness superimposed on a (N$_{\rm H}$, kT) grid
that is akin to an X-ray color-color diagram.

\subsection{Spitzer Observations}

We use infrared observations of the NGC 2071-IR region obtained by
the {\em Spitzer Space Telescope} to aid in X-ray source 
identification and to determine YSO classes
using mid-IR colors. {\em Spitzer}'s  Infrared Array Camera (IRAC) mapped
the NGC 2071-IR region as part of {\em Spitzer} Orion Survey  
(GTO program ID 043;  Megeath et al. 2005a; Allen et al. 2007; 
Flaherty \& Muzerolle 2008; Megeath et al. 2009, in prep.) 
Further information on IRAC can be found in Fazio et al. (2004).
The IRAC data were obtained in four 
channels (3.6, 4.5, 5.8, and 8.0 ~$\mu$m) using the high-dynamic 
range (HDR) mode, acquiring
a short 0.6 s exposure plus a longer 12 s exposure at each 
pointing position. Four dithers were obtained at
each position.  

The reduction and analysis of the complete
data set   will be  discussed in detail by Megeath et al. (2009, in prep.).
In brief, individual basic calibrated data (BCD) frames were
mosaicked with custom software to create separate mosaic images
for the four IRAC channels. Point sources were
identified using PhotVis v. 1.10 (Gutermuth et al. 2004;
Gutermuth et al. 2008).
Using the source positions determined from the mosaics,
photometry was measured for each point source on the
individual BCD images using the IDL module 
$aper.pro$  (Landsman  1993).  Magnitudes  were measured
in a circular aperture of radius  r = 2 pixels centered on the 
source, using  a background  annulus of r = 2 - 6 pixels.
A gain correction depending on pixel position was applied to each
magnitude. The final magnitudes are the median magnitude
returned from the four dithers. Magnitudes of
those sources in the NGC 2071-IR region discussed here
are given in Table 2, along with magnitude zero points and
aperture correction factors.

{\em Spitzer} MIPS data at longer wavelengths 24 - 160 $\mu$m are 
also available (Flaherty \& Muzerolle 2008; 
Megeath et al. 2009, in prep). 
Because  the NGC 2071-IR region is confused
and heavily  saturated,   we only provide here 
the MIPS 24 $\mu$m photometry for IRS 7.  MIPS
data were reduced using the procedure given in
Megeath et al. (2009).

\section{Results}         

We  focus here on the heavily-reddened NGC 2071-IR subgroup
shown in the 2MASS  2.16 $\mu$m (K$_{s}$) image in Figure 1 
and the {\em Spitzer} IRAC 3.6 $\mu$m (ch1) image in Figure 2. 
This region of high source density   includes the 
infrared sources IRS 1 - 8 and IRS 8a (W93; T07),
as well as the radio source VLA 1 located near IRS 1. 
We summarize the IR and X-ray properties  of each source
on an individual basis below.

Figure 3 shows near-IR colors of IRS 1-8, all of
which show some excess reddening. Figure 4
is an IRAC color-color diagram that 
provides one means of  distinguishing between 
class I YSOs (protostars) and class II YSOs
(classical T Tauri stars, or accreting star$+$disk systems).
Figure 5 plots the infrared spectral energy distributions
(SEDs) of those sources in NGC 2071-IR for which reliable
near-IR and {\em Spitzer} mid-IR data are available
(IRS 1, IRS 2, IRS 4, IRS 6, IRS 7).
Figure 6  shows the broad-band {\em Chandra} ACIS-I image for the
central region of NGC 2071-IR near IRS 1,
and Figure 7 is a hard-band  {\em Chandra} image
zoomed in on IRS 1 and VLA 1.
Figure 8 is a quantile diagram allowing the
X-ray  properties of {\em Chandra} detections to
be compared.

Table 1 summarizes the properties of the X-ray sources
detected by {\em Chandra}, and {\em Spitzer} IRAC
photometry is given in Table 2. {\em Chandra}
detected X-ray sources
at or near the positions of IRS 1,2,4,6 and VLA 1.
The infrared
sources IRS 5,7, 8, and 8a (offset $\approx$2$''$ northeast of IRS 8)
were undetected by {\em Chandra}.
Table 3 summarizes X-ray spectral analysis results
for the brighter detections.

\subsection{IRS 1}

IRS 1 was identified as a bright 10 $\mu$m source by
Persson et al. (1981).  It is also clearly visible in the 2MASS
K$_{s}$ image (Fig. 1) and in {\em Spitzer} IRAC images (Fig. 2).
A very bright source at or near the position of IRS 1 saturates
the  MIPS 24 $\mu$m images.
IRS 1 is likely the dominant contributor to the mid/far-IR luminosity
toward the NGC 2071-IR region, which was determined to be
L$_{tot}$ = 520 L$_{\odot}$ (d = 390 pc) by
Butner et al. (1990). This luminosity suggests that IRS 1 may be 
an embedded early-type star, as discussed further in
Section 5.1. But, its true nature is still somewhat of
a mystery.

IRS 1 is clearly detected in all four IRAC channels
and its SED rises sharply toward longer wavelengths
over the 3.6 $\mu$m - 8.0 $\mu$m IRAC range  (Fig. 5). 
It is the brightest IR source in the
NGC 2071-IR core region at  8.0 $\mu$m (Table 3).
{\em Spitzer}'s angular resolution of 
FWHM $\approx$ 1.7$''$ (3.6 $\mu$m) is 
unable to fully resolve VLA 1 located 
2.2$''$  north of IRS 1. But, the IRAC
images do show what appears to be faint
emission extending northward of IRS 1,
providing a clue that VLA 1 may be 
weakly contributing to the mid-IR flux
measured inside the r = 2.$''$4 aperture
centered on IRS 1.
Higher angular resolution mid-IR observations will 
be needed to confirm that VLA 1 is indeed a mid-IR
source.

The IRAC colors of IRS 1 are characteristic of a 
class I protostar  (Fig. 4).
Adopting the YSO classification scheme of Gutermuth
et al. (2008), class I sources are defined by
[4.5] $-$ [5.8] $>$ 1, where brackets denote
magnitudes at the enclosed  wavelength ($\mu$m).
IRS 1 has [4.5] $-$ [5.8] $=$
1.67 (Table 3), so the class I criterion is satisfied.
A similar conclusion is reached using the 
YSO classification schemes of Allen et al. (2004),
Hartmann et al. (2005), and Megeath et al. (2005b).

{\em Chandra} detects an X-ray source that is offset by 
only 0.42$''$ from the near-IR position of IRS 1 
(= 2MASS J054704.77$+$002142.8).
This offset is within {\em Chandra} positional uncertainties and
we thus associate X-ray source CXO J054704.75$+$002142.9 with 
the infrared source IRS 1. The IRAC 3.6 $\mu$m  detection of IRS 1  
has a centroid position J054704.76$+$002143.0, which is
offset by only 0.18$''$ from the {\em Chandra} X-ray position. 
This  {\em Chandra} detection confirms
that IRS 1 is an X-ray source, as suspected on the basis
of the previous lower spatial resolution {\em XMM-Newton} 
observation (S07).

The X-ray light curve of IRS 1 (Fig. 9) shows  
low-amplitude fluctuations about the mean at the $\pm$2$\sigma$ level
but no large flares. The KS test gives a probability of constant 
count rate P$_{\rm const}$ $=$ 0.11. Thus, low-level
variability may be present but is not proven at high confidence
levels.

IRS 1 is clearly detected in the hard-band ACIS-I image
(Fig. 7) but its emission is not as hard as that of 
VLA 1 or IRS 3 as gauged by median photon energy (Table 1).
The X-ray spectrum of IRS 1 (Fig. 10) is absorbed below 1 keV
but the absorption is not as high  as that of VLA 1. The ACIS-I
spectrum of IRS 1 shows an emission line from the Si XIII 
He-like triplet complex  (1.839 - 1.865 keV), implying thermal emission.
Maximum power for the Si XIII triplet is emitted at 
log T$_{max}$ = 7.0 (K) so hot plasma is clearly detected.
The spectrum also reveals a faint (7 counts) fluorescent emission line 
at  $\approx$6.4 keV from neutral or low ionization stages of
Fe (Fig. 11), as anticipated from {\em XMM-Newton} spectra. This line 
forms in ``cold'' material in the vicinity of the star that is 
irradiated by the hard X-ray source and is discussed further in Section 5.3.

We attempted to fit the  ACIS-I spectrum of IRS 1 with
various emission models. All models included
a fixed-width Gaussian line at 6.4 keV to reproduce the 
faint fluorescent Fe line. A  simple  one-temperature (1T)
APEC solar abundance optically thin plasma model 
underestimates the flux in the 1.7 - 1.9 keV range and 
is unacceptable in terms of $\chi^2$ statistics. 
This range includes the Si XIII He-like 
triplet lines at 1.839, 1.854, and 1.865 keV, which cannot
be individually distinguished at the spectral resolution 
of ACIS-I.  An acceptable 1T fit can be obtained by allowing 
Si to be overabundant by a factor of $\sim$3 with respect to 
its solar value. The apparent Si overabundance could either
be a real physical effect due to factors such as grain
destruction (e.g via grain-grain collisions in shocks)
or an  unphysical consequence of fitting the spectrum with an
overly simplistic 1T  model.

An acceptable solar abundance fit can be obtained with a
two-temperature (2T) APEC  model that includes both a cool and 
hot plasma  component (Table 3).  Some further improvement in 
the 2T fit  can be obtained by allowing Si to be overabundant,
but this is not required since the solar abundance
fit is already statistically acceptable ($\chi^2_{red}$ = 1.01).
Because of the heavy  absorption below 1 keV, the 
temperature and emission measure of any cool component are quite uncertain.
The 2T APEC model summarized in Table 3 uses a fixed temperature 
kT$_{1}$ = 0.55 keV for the cool component. This temperature results
in a minimum value of the C-statistic  (Cash 1979) when fitting
unbinned spectra. However, there is very little change in the
C-statistic for values in the range kT$_{1}$ = 0.25 - 0.70 keV.
The 2T model yields a best-fit temperature for the hot plasma component
kT$_{2}$ = 3.3 keV, but the true temperature could be higher since
the  90\% confidence upper limit is not tightly constrained by the data.
Using the conversion
N$_{\rm H}$  $=$ 2.22 $\times$ 10$^{21}$ A$_{V}$ cm$^{-2}$
(Gorenstein 1975), the solar abundance 2T APEC model
gives A$_{V}$ = 16 [12 - 20] mags, where square brackets
enclose the 90\% confidence range. The slightly different
conversion 
N$_{\rm H}$  $=$ 1.6 $\times$ 10$^{21}$ A$_{V}$ cm$^{-2}$
(Vuong et al. 2003) yields  A$_{V}$ = 22 [16 - 28] mags. 
The standard conversions used
above are reliable for low-extinction sources, but
some studies have questioned their accuracy for 
deeply-embedded YSOs (Winston et al. 2007).

The above APEC model assumes that the X-ray emission arises in
an optically thin thermal plasma and is typically used to model coronal
emission in magnetically active late-type stars. However,
other emission processes might be operating in IRS 1.
In particular, X-rays could be produced in shocks associated
with its powerful outflow. Thus, we tried to fit the
IRS 1 spectrum with an  isothermal plane-parallel
shock model (VPSHOCK). The fit statistic for  
VPSHOCK ($\chi^2_{red}$ = 1.16)  is not quite as good as 
the 2T APEC model and the fit converges to
shock temperatures kT $\approx$ 2 - 3 keV (Table 3).
Such shock temperatures are realistic but do require
high-velocity flows, as discussed further in 
Section 5.2.1.

In summary, a 1T APEC thermal plasma model is able to 
reproduce the {\em Chandra} spectrum of IRS 1 provided that  
Si is allowed to be overabundant. An acceptable solar-abundance 
fit can be obtained with a  2T APEC model, but it does not
tightly constrain the temperature or 
emission measure of the heavily-absorbed
cool component. This model requires hot plasma at 
kT$_{2}$ $\approx$ 3 keV and an absorption column density
equivalent to A$_{\rm V}$ $\approx$ 16 - 22  mags.

\subsection{VLA 1}

VLA 1 is a  compact  3.6 cm radio continuum source revealed
in high-resolution VLA A-configuration images obtained 
by  Carrasco-Gonz\'{a}lez et al. (2007). It was also marginally
detected in previous 5 GHz VLA images shown in Snell \& Bally (1986).
The 3.6 cm radio source is located 2.2$''$
north of IRS 1, which equates to a projected separation of
$\approx$860 AU at d = 390 pc.

The {\em Chandra} X-ray position is in very good agreement with
the  radio position of VLA 1 (Table 1). Due to its close proximity to IRS 1, 
VLA 1 was not spatially resolved from IRS 1 by
{\em XMM-Newton}. We are not aware of any published  high-resolution 
infrared images which show that VLA 1 is an infrared source.
But,  {\em Spitzer} images (Fig. 2) reveal
faint northward extension from IRS 1 that may be due to VLA 1.

The KS test shows that the X-ray emission of
VLA 1 is likely  variable with P$_{\rm const}$ $=$ 0.02, though the lightcurve 
shows no large amplitude flares (Fig. 9).
The ACIS-I spectrum  is heavily absorbed with little or
no X-ray emission detected below 2 keV. The spectrum can be 
acceptably fitted with a 1T APEC model at kT = 2.2 keV (Table 2). 
No significant Fe line emission is detected in the 6.4 - 6.7 keV
range. The inferred absorption column density log N$_{\rm H}$ = 22.94
gives an equivalent extinction A$_{V}$ = 39 [22 - 69] mag 
using the Gorenstein (1975) conversion. 
Based on X-ray spectral fits, VLA 1 is more heavily absorbed than
IRS 1 and Figure 8 indicates that it is the most heavily absorbed
X-ray source in NGC 2071-IR.  VLA 1 could thus be in an earlier 
evolutionary stage than IRS 1.
Because of the high absorption the intrinsic (unabsorbed)
X-ray luminosity of VLA 1 is quite uncertain.
But, the value inferred from
the 1T APEC model (Table 2) is consistent with that of a $\sim$solar-mass
pre-main sequence star or protostar, based on the known correlation between
L$_{\rm X}$ and stellar mass in low-mass YSOs
(Preibisch et al. 2005; Telleschi et al. 2007a).

\subsection{IRS 2}

Previous studies have concluded that IRS 2 is likely  a young 
stellar object (W93; T07) and Aspin et al. (1992)
argued that it is driving a bipolar outflow. IRAC images
show some indication that the source is extended.
IRAC colors (Fig. 4) indicate that IRS 2 is either a
class I protostar or a heavily-reddened class II object.
However, some caution is warranted in interpreting 
IRAC colors. High-resolution 3.6 cm radio continuum observations reveal
two closely-spaced  components A and B separated by 
$\approx$1.$''$4 with   the brighter component A at radio position
VLA J054705.367$+$002150.5
(Carrasco-Gonz\'{a}lez et al. 2009, private comm.).
{\em Spitzer} cannot resolve such a close pair and
it is thus possible that the IRAC colors are the
superposition of two sources.

{\em Chandra} detects a very faint X-ray source (4 net counts) at
an offset of only 0.27$''$ from the brighter radio component A (Table 1).
Because of the good positional coincidence with the radio source,
this weak X-ray detection is likely real but is of marginal significance.
There are too few
counts to do any substantive X-ray analysis but 
the high   median photon energy E$_{50}$ = 5.2  keV
suggests that this source is  exceptionally hard or
very heavily absorbed.

\subsection{IRS 3}

The infrared source IRS 3 is about one magnitude fainter
at K-band  and at 3.6 $\mu$m and 4.5 $\mu$m than IRS 1.
Its K band emission is highly 
polarized (W 93; T07). 
A 1.3 cm radio continuum source lies near IRS 3
(Torrelles et al. 1998), but the near-IR peak is
offset slightly to the northeast relative to the 
{\em VLA} radio  position (T07).
As Table 1 shows, the weak X-ray source detected by {\em Chandra}
is in excellent  positional agreement with the 1.3 cm radio 
source detected by Torrelles et al. (1998). The
nearest 2MASS source lies 2.3$''$ from the X-ray
peak. This offset is significant and supports the
conclusion of T07 that the
near-IR emission is nebulosity. 
The object detected with {\em Chandra} and the {\em VLA} is 
likely the star that illuminates the 
near-IR nebula. {\em Spitzer} IRAC images show an
infrared source near IRS 3 (Fig. 2). The IRAC position as measured
in the 4.5 $\mu$m (ch2) image is offset by only 0.42$''$  from 
{\em Chandra}. Thus, it appears that the self-luminous source  is
also detected by {\em Spitzer}. We do not have complete
IRAC photometry for IRS 3 because it lies in the
bright wings of IRS 1.

A KS test shows no significant X-ray 
variability. The distribution of event photon energies 
(Fig. 12) shows clearly that this is a hard absorbed
source, and the energy quantile plot (Fig. 8) 
substantiates this.  There were too few counts to analyze a binned 
spectrum but we did attempt to fit the unbinned spectrum with
a 1T APEC thermal plasma model using C-statistics (Cash 1979).
The absorption was held fixed at 
log N$_{\rm H}$ = 22.79 cm$^{-2}$, based on the estimated
extinction A$_{V}$ = 28 mag (W93). The fit converges to
a high but uncertain temperature kT $\gtsimeq$ 10 keV,
as also indicated by the quartile plot (Fig. 8). 

Visual inspection of the unbinned spectrum shows a weak buildup
of counts near 6.4 keV, hinting that a weak fluorescent Fe line
may be present. The IRS 3 event list reveals five
events within the 6 - 7 keV range with event energies 
6.11, 6.20, 6.42, 6.53, and 6.69 keV. The 6.42 and 6.53 keV
events could be due to fluorescent Fe.
We added a Gaussian line component
to the 1T APEC  model at a line energy of 6.4 keV,
fixing the line width at FWHM = 120 eV, corresponding to the 
ACIS-I instrumental width at this energy ({\em Chandra} POG, Fig. 6.8).
With the Gaussian line included, the C-statitistic C = 133.1 is
slightly less than without the line (C = 138.2), and the fit
gives  log N$_{\rm H}$ = 22.79 cm$^{-2}$,  kT = 10 keV, and an 
unabsorbed flux F$_{X,unabsorbed}$ (0.5 - 7.5 keV) =
1.6 $\times$ 10$^{-14}$ ergs cm$^{-2}$ s$^{-1}$.
This equates to log L$_{\rm X}$ = 29.5 ergs s$^{-1}$ (d = 390 pc),
which is typical of low-mass YSOs in Orion (Preibisch et al. 2005).
Thus, the event list and spectral fits provide some marginal
evidence for fluorescent Fe, but a deeper exposure would be
needed to make a compelling case for a 6.4 keV line detection.

The high-resolution {\em VLA} 1.3 cm radio continuum images of IRS 3
obtained by Torrelles et al. (1998) show elongation in the 
$\approx$N-S direction, and elongated morphology is also visible
in  the more recent 3.6 cm {\em VLA} images 
of Carrasco-Gonz\'{a}lez et al. (2007). The extension was interpreted
as a thermal radio jet, adding support to the belief that IRS 3 is
driving an outflow (Sec. 1). The orientation of H$_{2}$O maser spots 
approximately perpendicular to the jet axis and maser velocity
data led Torrelles et al. (1998) to conclude that the masers
trace a compact rotating molecular disk surrounding a low-mass
($\sim$1 M$_{\odot}$) YSO. Such a disk viewed in a nearly
edge-on geometry would act as a strong absorber of soft
X-ray photons and would provide a natural explanation for the high 
N$_{\rm H}$ (Fig. 8) and lack of detected low-energy X-ray photons      
below $\sim$2 keV in  IRS 3 (Fig. 12).

\subsection{IRS 4}

Tamura et al. (2007)  concluded  that IRS 4
is an IR cluster member because it is surrounded by
a compact reflection nebula   and is  associated with 
faint H$_{2}$ emission. 
The faint X-ray  source detected by {\em Chandra}
is in good positional agreement 
with 2MASS and shows no significant variability.
It has the lowest median and mean photon energy of the six X-ray
sources detected in the cluster core. There are insufficient
counts available for spectral analysis.
About half of the detected photons have energies below
2 keV. Quantile analysis  shows that this source
has lower  absorption than the other X-ray detections
in NGC 2071-IR (Fig. 8). Near-IR colors (Fig. 3) are consistent
with a star having intrinsic cTTS color but viewed through
subsantial reddening (A$_{\rm V}$ $\approx$ 13 mag).
IRAC colors (Fig. 4)  are consistent with a reddened 
class II YSO but  the SED (Fig. 5) is nearly flat 
in the IRAC bands and is more suggestive of a 
flat-spectrum source.

\subsection{IRS 5}

The near-IR morphology of this infrared source changes with time
and W93 concluded  that the near-IR emission could be from a jet,
a moving clump of shocked gas, or nebulosity. IRS 5 is located
between the two nearby bright mid-IR sources IRS 1 and IRS 2.
We were not able to identify a definite point source at the
IRS 5 position in the IRAC images, and thus cannot provide
any reliable IRAC photometry. No significant X-ray emission was 
detected so we find no evidence of a young star
and our results are thus consistent with the non-stellar
classification proposed by W93.

\subsection{IRS 6} 

Near-IR observations have shown that IRS 6 is  a binary 
with a 2$''$ separation (T07). We detect a variable X-ray source that
is offset to the northwest of the 2MASS position by 1.68$''$ 
at position angle PA = 315$^{\circ}$ (Table 1). This offset
is significant and implies  that the 2MASS near-IR 
source is not the same source detected in X-rays.
{\em Spitzer} images show an IRAC source at an
offset of 1.19$''$ from the {\em Chandra} position.
IRS 6 has similar near-IR and mid-IR colors  to
IRS 4, and both the near-IR and mid-IR colors of
IRS 6 are consistent with a reddened class II 
object (cTTS). 

The spectral fit of IRS 6 with a 1T APEC thermal model (Table 2)
suggests a rather hot source with plasma temperature kT = 3.85 keV
and moderate absorption N$_{\rm H}$ $\sim$ 10$^{22}$ cm$^{-2}$.
Similar values are inferred from the quantile plot (Fig. 8),
but both kT and N$_{\rm H}$ have rather large uncertainties due
to the low number of X-ray counts. 
The absorption from the  IRS 6 spectral fit (Table 2) gives an 
equivalent extinction 
A$_{V}$ = 5.6 [1.7 - 10.4, 90\% conf.] mag
(Gorenstein 1975 conversion), which  is less than IRS 1 and
VLA 1.

On the basis of its relatively high kT  and variability,
we conclude that the X-ray source lying 1.$''$68 from the
2MASS position is a  magnetically-active star. 
Because this system is a close near-IR pair,
higher spatial resolution observations are needed
to confirm the IR colors and more accurately determine the relative
positions of the two near-IR sources and the X-ray
source.

\subsection{IRS 7} 

Tamura et al. (2007) argued on the basis of
polarization angle that circumstellar material is 
present around IRS 7. IRAC colors support this conclusion, 
indicating  either a class I
protostar or a heavily-reddened class II source (Fig. 4).
The object is visible in MIPS 24 $\mu$m
images and the SED continues to rise out to
24 $\mu$m (Fig. 5). Its [4.5] $-$ [24] color is consistent
with a protostar (Megeath et al. 2009).
We detect no X-ray emission at or near IRS 7,
so it is either intrinsically X-ray faint or its
emission is reduced to undetectable levels by
high absorption (e.g. a nearly  edge-on disk or 
dense surrounding envelope).

\subsection{IRS 8 and 8a} 

IRS 8 and 8a are two faint closely-spaced IR sources
separated by 1.6$''$ in RA and 0.$''$7 in Dec, with
IRS 8a lying northeast of IRS 8 (W93; T07).
Both objects are $>$1.5 mag fainter at K  than the other
core sources (T07). 
The closest 2MASS positional match  to IRS 8 has K$_{s}$ = 13.54 
(2MASS J054704.45$+$002138.8) and there are no other
2MASS sources within 5$''$. 
IRS 8 is associated with a near-IR
nebula and was classified as a self-luminous
object, probably an embedded star, by  W93. IRS 8a
is thought to be a shocked molecular hydrogen
peak (W93).

The IRAC images reveal what appears to be extended
structure or nebulosity in the NE-SW direction near
IRS 8. A faint emission peak is visible close to the IRS 8
position at 3.6 $\mu$m (Table 2) and the  
peak shifts slightly to the northeast in the  
4.5 $\mu$m image, which is sensitive to shock emission.
The close pair IRS 8/8a is not clearly resolved at 
IRAC's resolution, but the slight shift of the 4.5 $\mu$m
emission peak toward IRS 8a  may be an indication that
it is a stronger shock-excited source. Because of the nebulous
morphology in IRAC images, we were not able to obtain
any reliable IRAC photometry. There is no X-ray emission
at or near the 2MASS position of IRS 8  or at IRS 8a,
and consequently no X-ray evidence of an embedded star.

\section{Comparison with XMM-Newton Results}

The {\em XMM-Newton} EPIC CCD spectrum of IRS 1 analyzed by 
S07 was extracted using a nominal circular region of radius 
R$_{e}$ = 15$''$ (68\% encircled energy) centered on
IRS 1. As the higher angular resolution {\em Chandra}
image in Figure 6 shows, four X-ray sources were 
included in the {\em XMM-Newton} spectral extraction
region, namely IRS 1,2,3 and VLA 1. Of these four,
IRS 1 (144 counts) and VLA 1 (90 counts) are  the 
brightest NGC 2071-IR core objects in the  {\em Chandra} image.
IRS 2 (4 counts) is very faint in  ACIS-I
and it is unlikely that it significantly
affected the  {\em XMM-Newton} spectrum.
IRS 3 (18 counts) is also a faint hard {\em Chandra}
source. Its absorbed
flux  F$_{\rm X}$ (0.5 - 7.5 keV) =
8 $\times$ 10$^{-15}$ ergs cm$^{-2}$ s$^{-1}$ would
have contributed $\approx$5\% to the observed
flux measured in the {\em XMM-Newton} pn spectrum.

Based on the above, it is clear that the {\em XMM-Newton} 
spectrum was dominated by emission from the close pair
IRS 1 and VLA 1. The {\em Chandra} data show that
VLA 1  has much higher X-ray absorption than IRS 1,
which explains why fits of the  {\em XMM-Newton} 
pn spectrum mysteriously required two spectral components with
different N$_{\rm H}$ values (S07).

The fluorescent Fe line at 6.4 keV was strongly detected in
the {\em XMM} pn spectrum with an EPIC pn  line flux
F$_{\rm X}$ (6.2 - 6.6 keV) =
3.8 ($\pm$0.2) $\times$ 10$^{-14}$ ergs cm$^{-2}$ s$^{-1}$ 
and an equivalent width (EW) measured relative to the
underlying continuum EW = 2.4 keV (S07). The line
flux measured in the {\em Chandra} ACIS-I spectrum
(Table 3) is a factor of $\sim$5 below the {\em XMM}
value, but the {\em Chandra} EW $\geq$ 2.4 keV is consistent 
with the {\em XMM} value. The fluorescent Fe line is discussed
further in Section 5.3.

\section{Discussion}

\subsection{Is IRS 1 a Young  Embedded B-type Star?}

The heavily-reddened IRAC colors of IRS 1 and its
rising SED in the IRAC bands, along with  its bright (saturated)
detection in MIPS 24 $\mu$m images,  strongly
suggest that it is a class I protostar. Based
on its association with a radio continuum source
and luminosity considerations, W93 concluded that
IRS 1 is most likely a B0-B5 star.
Under the assumption that the 6 cm
radio continuum emission of IRS 1 arises in a photoionized
compact H II region at a distance of 500 pc, 
Snell \& Bally (1986) 
found that a single B2 star  would be 
needed to produce sufficient  Lyman continuum ionizing 
flux to account for the radio emission. Adjusting their
distance downward to 390 pc (Anthony-Twarog 1982)
would require a $\sim$B3 star.
However, we note that the volume emission measure
of the radio continuum source computed from their 
6 cm radio flux density S$_{6cm}$ = 7.9 mJy is 
two orders of magnitude below that  typically
found for optically thin compact HII regions 
(eq. [3] of Skinner, Brown, \& Stewart 1993). 
Thus, the radio continuum emission from IRS 1
could very well involve processes other than
free-free emission from a compact HII region.

The more recent high-resolution 3.6 cm {\em VLA}
A-configuration images of IRS 1 obtained by
Carrasco-Gonz\'{a}lez et al. (2007) reveal a
complex source morphology for IRS 1. A compact
core is present along with fainter emission 
extending in the E-W direction. More interestingly,
compact knots are seen expanding away from IRS 1
in an approximate E-W direction with high velocities
of up to $\sim$600 km s$^{-1}$, as determined  from 
radio proper motions. These results, along with the
positive radio spectral index,  suggest that the radio 
emission includes a contribution from an extended ionized
wind or jet. A possible analog is  NGC 7538-IRS1,
whose free-free emission is dominated by a highly
collimated ionized jet (Sandell et al. 2009).

Using far-IR data obtained with the {\em Kuiper Airborne 
Observatory} (KAO), Butner et al. (1990) determined the
total luminosity toward the NGC 2071-IR region to be
L$_{tot}$ = 520 L$_{\odot}$ (d = 390 pc). This is
about 15\% greater than the value determined by
Harvey et al. (1979) when normalized to a common
distance of 390 pc.  Because of
the rather large $\approx$30$''$ KAO beam size, this
total will  include any contributions from other sources
near IRS 1 and should thus be considered an upper limit
for IRS 1 itself. If due to a single star, a value
L$_{tot}$ =  520 L$_{\odot}$ would correspond to a 
ZAMS star of mass $\sim$5 M$_{\odot}$ (Siess, Dufour, \&
Forestini 2000) and ZAMS spectral type $\sim$B4 (Thompson 1984). 
But, if the L$_{tot}$ value is interpreted as an upper 
limit for IRS 1, a lower mass and later spectral 
type are possible. The above estimates are based on
an assumed distance d = 390 pc. If IRS 1 lies at
the far edge of the Orion OB1 association then
a larger distance d $\sim$ 500 pc is possible (Sec. 1).
In that case, one infers a slightly higher mass
of $\sim$6 M$_{\odot}$ and  ZAMS spectral
type of $\sim$B3 - B4.

The X-ray luminosity and temperature of IRS 1 are within
the broad range observed in young B-type stars. 
The study of 20 early-type stars in
the {\em Chandra} Orion COUP sample by
Stelzer et al. (2005) included 
12 B stars ranging in spectral type from
B0.5 to B9. Their X-ray luminosities spanned a
wide range from log L$_{\rm X}$ = 29.4 - 32.4
ergs s$^{-1}$, and a hot plasma component at
typical  temperatures  kT$_{hot}$ $\approx$ 
1 - 3 keV was present in all B-stars. 
Different B  stars with similar
spectral types showed dramatic differences in
L$_{\rm X}$ and no support was found for earlier
claims that L$_{\rm X}$ correlates with L$_{bol}$
in OB stars.  Using L$_{bol}$ = 520 L$_{\odot}$
for IRS 1 and the range of L$_{\rm X}$
values from the different models in Table 3,
we obtain log (L$_{\rm X}$/L$_{bol}$) =
$-$5.7 $\pm$0.5. This is consistent with 
values obtained for the COUP B-star sample,
but some caution is needed in interpreting 
this ratio for IRS 1 since its L$_{bol}$
is could be  dominated by disk emission.

On the basis of the above comparisons, we conclude
that a B-star classification for IRS 1 is plausible,
but a mid-to-late B star is more likely than an
early B star based on the  luminosity determined 
by Butner et al. (1990). If IRS 1 is indeed an 
embedded intermediate-mass B-star, it is tempting
to speculate that it might be a Herbig Ae/Be star
progenitor. Herbig Ae/Be stars are optically-revealed
intermediate-mass pre-main sequence stars which in
many cases show high-temperature X-ray plasma 
that is likely of magnetic origin. But, the question
of whether their X-ray emission is intrinsic or
instead due to coronal late-type companions 
remains open.
For previous studies of X-ray emission from Herbig
Ae/Be stars, the reader is referred to 
Damiani et al. 1994; Zinnecker \& Preibisch 1994; 
Preibisch \&  Zinnecker 1996; Skinner \& Yamauchi 1996; 
Skinner et al. 2004; Hamaguchi, Yamauchi, \& 
Koyama 2005; Stelzer et al. 2006, 2009; 
and Telleschi et al. 2007b.

\subsection{X-ray Emission Processes in IRS 1}

Given that the true nature of  IRS 1  is not
well-known and that it may be an embedded
B-type star, some  further discussion of the origin of
its X-ray emission is warranted. Very few massive
embedded stars have been detected in X-rays and little
is known about their X-ray emission  during early evolutionary
stages prior to becoming optically visible. We thus consider
several possible emission mechanisms below.

\subsubsection{Shocked Winds and Outflows}

The  hotter plasma clearly detected at 
kT$_{hot}$ $\gtsimeq$ 2 keV is not consistent with 
the cool emission (kT $<<$ 1 keV) expected from wind 
shocks in OB stars formed by line-driven flow instabilities
(Lucy \& White 1980; Lucy 1982; Feldmeier et al. 1997; 
Owocki et al. 1988). However, if the wind is shocking
onto another object or  surrounding material, or
is magnetically-confined, then high-temperature plasma
could be produced in a wind shock.

The predicted shock temperature for a strong adiabatic
shock is
T$_{shock}$ = 1.4 $\times$ 10$^{5}$~$\Delta$v$_{100}^2$~K,
or kT$_{shock}$ = 0.012$\Delta$v$_{100}^2$ keV
(Lamers \& Cassinelli 1999). Here,
$\Delta$v$_{100}$ is the shock jump  measured 
relative to the speed of the downstream flow, in units of
100 km s$^{-1}$.
To achieve an  X-ray temperature kT$_{shock}$ $\sim$ 2 keV,
a  value $\Delta$v $\sim$ 1300 km s$^{-1}$ is required.
Terminal wind
speeds of B5 - B9 stars  are not well-determined
observationally, but values
v$_{\infty}$ $\sim$ 1200 - 1400 km s$^{-1}$ are usually assumed
(Cohen et al. 1997). To achieve the required 
velocity jump, the wind would have to be virtually
stopped by dense surrounding material, e.g. either
a disk, thick envelope, or close companion. 
The wind speed  requirement
could be reduced if the outflowing wind collides
with a dense infalling envelope whose ram pressure
exceeds that of the wind. In that case, the infalling
envelope would overpower the wind, resulting in a
non-stationary shock front (including both forward
and reverse shocks) moving toward the star.

The above analysis can also be applied to jets
and molecular outflows. 
High velocity wings extending out to 
$\approx$70 km s$^{-1}$ are visible in
spectra of the 2.12 $\mu$m H$_{2}$ line
toward NGC 2071-IR (Persson et al. 1981).
If such an outflow shocks onto a stationary
target, the maximum expected X-ray temperature
is kT$_{shock}$ = 0.006 keV, much
too low to explain the hot X-ray plasma detected
in IRS 1. As already noted (Sec. 5.1), high-velocity
knots moving outward from IRS 1 at speeds up to
$\sim$600 km s$^{-1}$ have been detected with 
the {\em VLA} (Carrasco-Gonz\'{a}lez et al. 2007).
If such a knot shocked onto a stationary 
object, the maximum shock temperature would be
kT$_{shock}$  $\approx$ 0.4 keV, which  is again too
low to explain the higher temperature plasma in
IRS 1. But, this process could contribute to 
any cooler heavily-absorbed emission at 
kT $<$ 1 keV.

\subsubsection{Accretion Shocks}

Since IRS 1 is a young class I object, it is  
expected to still be accreting. This is potentially
relevant to the X-ray interpretation because  X-rays 
can be produced in an accretion shock formed as
infalling material impacts the   (proto)star.
The characteristic
X-ray temperature for an accretion shock is
kT$_{acc}$ $\approx$ 0.02v$_{\rm 100}^2$ keV,
where  v$_{\rm 100}$ is the infall speed in units of 
100 km s$^{-1}$ (Ulrich 1976). 

Based on the above, it is apparent 
that very high infall speeds   
v $\sim$ 1000 km s$^{-1}$ would be needed 
to  produce accretion shock temperatures 
kT$_{acc}$ $\sim$ 2 keV,  comparable to the
values  observed for IRS 1. To achieve such
infall speeds under free-fall
conditions,  a massive
central object equivalent to a B0V star
(M$_{*}$ $\sim$ 18 M$_{\odot}$,
R$_{*}$ $\sim$ 7 R$_{\odot}$; Allen 1985)
would be required.
But, the total luminosity of IRS 1
(Sec. 3.1) falls well short of the value
L$_{bol}$ $\sim$ 10$^{4.4}$ L$_{\odot}$
expected for a B0 ZAMS star. Thus, in the absence of
any firm observational evidence for very high
infall speeds toward IRS 1, it is very unlikely
that the high-temperature X-ray emission
arises in an accretion shock. Any cooler
plasma (kT $<$ 1 keV) that might be present
could be accretion-related but, as already
noted, the high absorption toward IRS 1
masks such cool emission.

\subsubsection{Magnetic Activity}

Plasma temperatures kT $\gtsimeq$ 2 keV as seen in
IRS 1 are commonly found in magnetically-active
late-type stars, including T Tauri stars.
The X-ray emission is accompanied by powerful
magnetic reconnection flares, and is thought
to arise in structures analogous to the solar
corona.  But, magnetic star$+$disk coupling 
undoubtedly leads to more  complicated geometries 
in the case of TTS (reviewed by Feigelson \& Montmerle 1999).
Similarly, high plasma temperatures and large
flares characteristic of magnetospheric processes
have now been observed in low-mass 
class I protostars such as those  in 
$\rho$ Ophiuchus (Imanishi, Koyama, \& Tsuboi 2001).
X-ray luminosities in
low-mass YSOs generally increase with stellar mass 
and luminosity,
ranging from L$_{\rm X}$ $\sim$ 10$^{28}$ ergs s$^{-1}$
for low  masses M$_{*}$ = 0.1 - 0.2 M$_{\odot}$ up to
(and  in some cases above)  L$_{\rm X}$ $\sim$ 10$^{31}$ ergs s$^{-1}$  
for more massive TTS with M$_{*}$ $\approx$ 2 M$_{\odot}$
(Preibisch et al. 2005; Telleschi et al. 2007a).

The X-ray temperature and luminosity of IRS 1
inferred from the {\em Chandra} data are well
within the above ranges for low-mass YSOs.
It is thus possible that the X-rays observed 
toward IRS 1 arise in a low-mass YSO rather than
a more massive object such as an embedded B-type star. 
The X-ray luminosity of IRS 1 could be accounted for
by a  TTS of mass $\sim$1 - 2 M$_{\odot}$.
Even if IRS 1 is an embedded B-type star, a
late-type TTS companion could  dominate the 
X-ray emission. Late-type companions have
been proposed as a means of explaining 
the wide range of L$_{\rm X}$ values observed
for some mid-to-late B stars in Orion with weak winds
(Stelzer et al. 2005)

\subsubsection{Magnetically-Confined Wind Shocks}

If IRS 1 has an ionized wind and a sufficiently
strong magnetic field,
then the B field can confine
the wind into two oppositely directed streams in
each hemisphere which collide near the magnetic
equator, forming a magnetically-confined wind
shock (MCWS). Such a shock can reach X-ray
emitting temperatures.  
The MCWS  mechanism has been used to
explain high-temperature X-ray emission in 
magnetic Ap-Bp stars (Babel \& Montmerle 1997a)
and young magnetic O stars such as 
$\theta^1$ Ori C (Babel \& Montmerle 1997b; 
Gagn\'{e} et al. 2005).

The radio properties of 
of IRS 1 do  suggest it has an ionized wind or
jet (Carrasco-Gonzalez et al. 2007), but there are so 
far no reports of a  magnetic field. Even so, an 
estimate of the field strength
needed to confine the wind can be obtained
by making reasonable approximations under
the assumption that it is a mid-to-late B star.

The maximum X-ray temperature for a MCWS is given
by the adiabatic shock  equation (Sec. 5.2.1).
We thus  assume a terminal wind speed 
v$_{\infty}$ $\sim$ 1300 km s$^{-1}$ for IRS 1
to achieve a shock temperature comparable to 
the observed value (kT$_{shock}$ $\sim$ 2 keV).
The degree to which the wind is confined by the
magnetic field is expressed in terms of the 
confinement parameter 
$\eta$ = B$^2$R$_{*}^2$/$\dot{M}$v$_{\infty}$
where B is the equatorial  magnetic field 
strength  (ud-Doula \& Owocki 2002).
For values of $\eta$ $\approx$ 1 (critical confinement),
the B field is able to confine the wind.
For a B5V star (R$_{*}$ $\approx$ 3.8 R$_{\odot}$,
Allen 1985), the above relation with $\eta$ = 1 becomes
B$_{100}^2$ $\approx$ 47$\eta$$\dot{M}_{-6}$,
where B$_{100}$ is in units of 100 G
and $\dot{M}_{-6}$ is in units of 
10$^{-6}$ M$_{\odot}$ yr$^{-1}$.
The mass-loss rates of mid-to-late B
stars are not well-determined empirically,
but B $\sim$ 2 G for $\dot{M}$ =
10$^{-11}$ M$_{\odot}$ yr$^{-1}$ 
(Cohen et al. 1997) and 
B $\sim$ 70 G for 
$\dot{M}$ = 10$^{-8}$ M$_{\odot}$ yr$^{-1}$.
The latter $\dot{M}$  values are typical of 
intermediate mass 
Herbig Ae/Be stars (Skinner et al. 1993).

In summary, the MCWS mechanism is a potential 
means of explaining the high-temperature plasma in
IRS 1. But, without a magnetic field detection
and specific estimates of B-field strength and
mass-loss parameters, any more definitive comparisons against
MCWS model predictions cannot be made.

\subsubsection{Summary of Emission Mechanisms}

Based on the above discussion, we conclude that the 
high-temperature plasma at kT $\sim$ 2 keV in IRS 1
is most likely due to either a high-speed wind
(v$_{\infty}$ $\gtsimeq$ 1200 km s$^{-1}$) shocking onto a 
dense  target (e.g. a disk or dense
surrounding envelope), or it originates in a
low-mass magnetically-active YSO companion. The wind-shock
interpretation is plausible if IRS 1 is an embedded B-type
star whose wind has already turned on, and
the extended  bright radio continuum emission detected
with the {\em VLA} does point to an early-type star
with a strong wind or jet. We cannot rule out a
magnetically-confined wind shock, but there are
insufficient observational constraints on magnetic
field strength and wind parameters to rigorously
test MCWS models. Even if IRS 1 is an embedded 
B-type star (which seems likely), a lower mass
YSO companion at close separation  could still
be responsible for some (or even all) of the observed
X-ray emission.   The X-ray luminosity determined
from spectral  models of IRS 1 is at the high end
of the range observed for low-mass YSOs.  If the
X-ray emission is due entirely to a T Tauri-like companion,
then the known correlation between L$_{\rm X}$ and stellar 
mass suggests a companion mass in excess of 1 M$_{\odot}$.

\subsection{Fluorescent Iron Emission in IRS 1}

A fluorescent Fe line at 6.4 keV can be produced by
photoionization when ``cold'' material containing neutral 
or weakly-ionized iron is irradiated by a nearby hard X-ray 
continuum source. Photons with energies above 7.11 keV
are needed to eject a K-shell electron.
The K-shell vacancy is filled by another electron
such as an L-shell electron, resulting in 
fluorescent Fe line emission at 6.4 keV. The
line is in fact a doublet consisting of two
closely-spaced lines at 6.391 keV and  6.404 keV,
but  these cannot be distinguished at the energy 
resolution of the ACIS-I CCDs. The above photoionization
process is discussed in more detail by 
George \& Fabian (1991) and Kallman et al. (2004).
Analysis of the fluorescent Fe line can potentially
provide information on the properties of the
absorber and its location relative to the 
hard continuum source. 

The fluorescent Fe line detected in IRS 1 
is faint (Fig. 11),
consisting of 7 events distributed in the 
6.298 - 6.536 keV energy range, with a median
energy 6.395 keV and mean energy 6.419 keV.
Three of the seven photons arrived in the first
half of the observation. 
A KS test applied to the arrival times of
the 7 events shows no significant variability
and gives a probability of constant count rate
P$_{const}$ = 0.35. The line flux is 
7.2 ($\pm$1.2) $\times$ 10$^{-15}$ ergs cm$^{-2}$ s$^{-1}$.
This is about a factor of five less than measured
during the March 2005 {\em XMM} observation (S07).
The line equivalent width in the ACIS-I spectrum
is uncertain because the continuum flux near
6.4 keV is very faint. But, our flux estimates
give values EW $\geq$ 2.4 keV, consistent with
the previous {\em XMM} estimate (S07).

The detection of fluorescent Fe line emission 
in IRS 1 is a rare occurrence in YSOs, but 
certainly not unique.
Tsujimoto et al. (2005) examined 1616 X-ray sources
detected in the deep {\em Chandra} COUP observation
of the Orion Nebula Cluster and identified seven
sources with excess emission at 6.4 keV and 
line equivalent widths EW $\leq$ 0.27 keV.
All seven X-ray sources showed flare-like variability
and high line-of-sight absorption 
N$_{\rm H}$ $\gtsimeq$ 10$^{22}$ cm$^{-2}$,
as also seen in IRS 1. The broad fluorescent Fe line during a
superhot flare in V1486 Ori (COUP source 331) was  analyzed by
Czesla \& Schmitt (2007). The 6.4 keV line has also
been detected in other YSOs including the class I objects
YLW 16A (Imanishi et al. 2001) and Elias 29 (Giardino et al. 2007)
in $\rho$ Oph, and the class I source  R CrA X$_{E}$
(= IRS 7B; Hamaguchi et al. 2005).

The fluorescent Fe line detection in IRS 1 is 
interesting in three respects:
(i) it was detected  in
the apparent absence of any large-amplitude flares  
(but low-level variability may be present),
(ii) there is no accompanying line at 6.7 keV 
from the highly-ionized Fe XXV complex, and
(iii) the line has a very large equivalent width
that is difficult to account for by illuminated
disk models.

The absence of any detectable  large flares in IRS 1 is 
unusual because the fluorescent Fe lines
in almost all of the YSOs mentioned above 
occurred during flares. Analysis of the 
V1486 Ori data showed that its broad fluorescent
Fe line appeared during the rise phase of a
flare (Czesla \& Schmitt 2007). An interesting 
exception is Elias 29, which showed a 
6.4 keV line  during a large flare. But the line
persisted after the flare had decayed, remaining
over a timescale much longer than the radiative
lifetimes of  fluorescent Fe transitions. Giardino
et al. (2007) thus concluded that the X-ray flare
was not directly responsible  the post-flare 6.4 keV 
line. Instead, they proposed that the 
line was produced by collisional ionization 
from an unseen population of nonthermal electrons.

The 6.4 keV line in IRS 1 is similar to Elias 29 in
the sense that it does not appear to be directly
associated with a large flare. The line was  present 
in both the {\em XMM-Newton} spectrum obtained
in March 2005 and in the more recent {\em Chandra}
spectrum, but no large flare was detected in either
observation. If a large flare was involved, then it
must have occurred prior to the start of the observations,
or it escaped detection (as could  occur if the flare
occurred on the back side of IRS 1 and was occulted).

Unless large X-ray flares occurred and escaped detection,
the existing data  indicate that the 6.4 keV line
in IRS 1 is persistent and does not correlate with large flares.
In that case, some  mechanism needs to be operating nearly
continuously to ionize the absorbing material. At
present, the ionizing source and the ionization mechanism
are not known. The obvious candidate for the ionizing
source is IRS 1 itself. But, in terms of hardness, its
{\em Chandra} spectrum is rather benign,  showing no 
large flares, no Fe XXV line and a weak continuum
above $\sim$7 keV.  However, we cannot rule out the
possibility of  substantial hard emission 
(kT $\gtsimeq$ 20 keV) above the energy  range accessible to 
{\em Chandra} and {\em XMM}. Nevertheless, the intriguing
possibility does remain that the hard ionizing source is 
not IRS 1 and is not directly  detected in our 
X-ray observations.  

However, before discounting IRS 1 as the ionizing source,
two factors should be noted. First, low-level variations
appear to be present in  the {\em Chandra}   
light curve of IRS 1 (Fig. 9). These variations
may be a signature of persistent low-level flaring.
If that is the case, then quasi-continuous hard energy
release by repeated small flares may play a role in
photoionizing the absorber. Second, IRS 1 is thought
to drive a powerful outflow and  also shows evidence of 
a radio jet (Carrasco-Gonz\'{a}lez et al. 2007).
If a population of nonthermal electrons is produced
in the shocked jet (which may be magnetically collimated)
they could play a 
role in collisional ionization of the absorber.
Collisional ionization has been invoked as a means
of explaining the 6.4 keV  line in Elias 29
(Giardino et al. 2007) and in the active binary
system II Peg (Osten et al. 2007). 
If a population of nonthermal paricles is present, they {\em may}
imprint a signature on the radio continuum emission. Thus,
further radio observations of IRS 1 at lower frequencies where 
nonthermal emission can dominate over thermal (free-free) emission
could be informative.

The  large equivalent width EW $\geq$ 2.4 keV of the IRS 1 
fluorescent line is a noteworthy feature of its X-ray spectrum.
This value is extreme for a YSO, exceeding even the
maximum value EW = 1.4 keV
observed for V1486 Ori during the rise phase of its superhot flare
(Czesla \& Schmitt 2007). These values conflict with theoretical
models of centrally illuminated cold disks, which predict
EW $\approx$ 0.1 - 0.2 keV (George \& Fabian 1991).
The EW of the 6.4 keV line is
proportional to the column density N$_{\rm H'}$ 
in the fluorescing material (Tsujimoto et al. 2005). 
If taken literally, the 
large EW of IRS 1 would imply a high absorber
column density N$_{\rm H'}$ $\gtsimeq$ 10$^{24}$ cm$^{-2}$
(S07). Since this value is much larger than the
line-of-sight photoelectric absorption inferred from
the X-ray spectrum (Table 3), one is led to the conclusion
that the  fluorescing material lies off the line-of-sight. 
But, the above conclusion rests on the assumption that
the continuum measured in the X-ray spectrum has 
undergone negligible absorption. If the hard irradiating
source is not IRS 1 itself and is occulted or very 
heavily absorbed, then some or all of the  intrinsic 
hard continuum will be
missing from the observed X-ray spectrum. In that case,
the EW measurement would be based on a depressed continuum and
thus overestimated, along with  N$_{\rm H'}$.

In summary, our main conclusions with respect to the fluorescent
Fe line in IRS 1 are: (i) the line is present even in the 
absence of the Fe XXV line (6.67 keV) and any large flare 
detections, and evidently does not directly depend on 
large-amplitude X-ray flares for its existence,
(ii) the line flux is variable, showing a decrease  between the 
{\em XMM} and {\em Chandra} observations taken 2.6 years
apart,  but no significant variability during the 67 ksec 
{\em Chandra} observation, (iii) the line EW is much larger
than predicted by centrally-illuminated disk models, 
suggesting either that the photoionization models do not 
accurately reflect the line formation process and that 
other mechanisms (e.g. collisional ionization) could play 
a role, or that the observed continuum is not the same 
as the intrinsic continuum of the photoionizing source
(e.g. due to the effects of absorption or occultation),
and (iv) the hard ionizing source may have so far escaped 
X-ray detection.

\section{Conclusions}

The main conclusions of this study are the following:

\begin{enumerate}

\item {\em Chandra}'s excellent spatial resolution has 
      resolved the  NGC 2071-IR core region into six
      X-ray sources. X-ray emission was detected at
      or near the infrared sources IRS 1,2,3,4 and 6 
      and the radio source VLA 1. X-ray analysis shows
      that these sources are viewed through moderate-to-high
      absorption and have hard/high-temperature X-ray
      emission (kT $\gtsimeq$ 2 keV), consistent with
      their classification as embedded young stars.
      In some cases such as IRS 3, the X-ray position
      is significantly offset from the near-IR position, 
      supporting the conclusion that the near-IR peak
      traces  nebulosity or scattered light and not the
      star itself.

 \item {\em Spitzer} IRAC detections of IRS 1,2,4,6, and
       7 provide the first reliable mid-IR colors for these
       objects. Using IRAC colors along with near-IR colors
       based on published JHK$_{s}$ photometry and the sharply-rising
       IR SED, we conclude that
       IRS 1 is a class I protostar. IRS 2 and IRS 7 are also
       likely class I protostars, but could be heavily-reddened
       class II objects. IRS 4 and IRS 6 have colors consistent
       with reddened class II objects.  No reliable 
       IRAC photometry or mid-IR colors were obtained for 
       IRS 3, 5, 8, or 8a, but near-IR colors determined from
       existing JHK$_{s}$ photometry  show that these objects are 
       heavily-reddened.

  \item {\em Chandra} provides the first X-ray detection of
        VLA 1, and its X-ray emission is likely  variable.  
        X-ray spectral models require a strong absorption 
        component for VLA 1, equivalent to  
        A$_{\rm V}$ = 39 [22 - 69] mag. Thus, VLA 1 appears
        to be more heavily-embedded than IRS 1 and it could thus be
        an even younger, less-evolved object.
        There are as yet no IR data for VLA 1 on which to base a 
        YSO classification.

\item   The existing data for IRS 1 are consistent with a 
        mid-to-late B (proto)star classification.
        The X-ray spectrum of IRS 1 is heavily-absorbed below 
        1 keV and clearly reveals high-temperature plasma 
        at kT $\sim$ 2 - 3 keV. Such plasma could originate either
        in a shocked high-velocity wind/jet or in a magnetically-active
        low-mass YSO companion. 

\item The most important finding of this study  is that the 
      fluorescent Fe line  in IRS 1 is persistent and  present
      even in the absence of large flares.  Thus, an as yet unidentified
      quasi-continuous process that operates independently of large
      flares is needed to explain the K-shell ionization of cold
      proximate material that leads to formation of fluorescent 
      Fe.  Both the {\em Chandra} and {\em XMM-Newton} light
      curves show signs of low-level variability, suggesting that
      short bursts of hard emission from persistent low-level flaring  
      could play a role in ionizing the absorber. The large
      equivalent width EW $\geq$ 2.4 keV of the 6.4 keV line
      is noteworthy because it is well in excess of values predicted
      by centrally illuminated disks models. Large 6.4 keV line 
      equivalent  widths have also been noted in other YSOs 
      such as V1486 Ori. If interpreted
      literally, the large EW values imply either a very high
      absorption column density in the cold absorber, or perhaps
      processes other than photoionization (e.g. collisional
      ionization in shocks).  However, some
      caution is needed in interpreting the large EW for IRS 1,
      since the absence of large flares raises the possibility
      that it is not the hard ionizing source. If the ionizing
      source is occulted by IRS 1 or buried behind heavy
      absorption, then the hard continuum measured in the
      IRS 1 spectrum may be less than the intrinsic continuum
      of the  ionizing source (``missing continuum''), leading
      to an overestimate of the equivalent width.

\end{enumerate}

\acknowledgments

S.S., K.F., and M.M. acknowledge {\em Chandra} support from SAO
grant GO7-8008A.
M.A. acknowledges support from a Swiss National Science 
Foundation Professorship (PP002--110504).
We have utilized data products from the Two Micron All-Sky Survey
(2MASS), which is a joint project of the University of Massachusetts
and IPAC/CalTech. This work is based in part on  data
obtained with the {\em Spitzer} Space Telescope, which is
operated by the Jet Propulsion Laboratory, CalTech, under a
contract with NASA. We would like to thank  
C. Carrasco-Gonz\'{a}lez and G. Sandell   for 
radio positions and J. Hong for information on
quartile analysis computer codes.

\clearpage

\begin{deluxetable}{lllllllll}
\tabletypesize{\scriptsize}
\tablewidth{0pt} 
\tablecaption{X-ray Sources Near NGC 2071-IRS 1\tablenotemark{a}}
\tablehead{
	 \colhead{Name}                                    &
           \colhead{R.A.}               &
           \colhead{Decl.}                             &
           \colhead{Net Counts}                   &
           \colhead{E$_{25}$,E$_{50}$,E$_{75}$}                                       &
           \colhead{P$_{\rm const}$}                                     &
           \colhead{K$_{s}$}                   &
           \colhead{Identification(offset)}                                    \\
           \colhead{}                                          &        
           \colhead{(J2000)}                         &
           \colhead{(J2000)} &
           \colhead{(cts)}                                          &               
           \colhead{(keV)}                                          &            
           \colhead{}                                          & 
           \colhead{(mag)}                                          & 
           \colhead{(arcsec)} 
                                  }
\startdata
IRS 3\tablenotemark{b} & 05 47 04.63 & +00 21 48.0 &  18$\pm$4     & 3.68,4.52,6.06 & 0.76 & 12.18\tablenotemark{c} & VLA J054704.62+002147.8 (0.25)\tablenotemark{d}\\
IRS 1 & 05 47 04.75 & +00 21 42.9 & 144$\pm$12    & 1.79,2.22,3.27 & 0.11 & 11.21  & 2M J05470477+0021428 (0.42)\\
VLA 1 & 05 47 04.77 & +00 21 45.1 &  90$\pm$10(v) & 2.98,3.52,4.48 & 0.02 & ...    & VLA J054704.758+002145.50 (0.44)\tablenotemark{e}\\
IRS 4 & 05 47 05.10 & +00 22 01.6 &  11$\pm$3     & 1.18,1.75,2.53 & 0.14 & 10.86  & 2M J05470512+0022013 (0.48) \\
IRS 2 & 05 47 05.35 & +00 21 50.4 &   4$\pm$1\tablenotemark{f}   &...~~~,5.22,~...  & ...\tablenotemark{f}  & 10.41 & 2M J05470538+0021500 (0.66)\tablenotemark{e}\\
IRS 6 & 05 47 05.60 & +00 22 11.1 &  48$\pm$7(v)  & 1.56,2.05,3.21 & 0.01 & 10.10  & 2M J05470570+0022103 (1.68)\tablenotemark{g} \\
\enddata
\tablenotetext{a}{
Notes:~{\em Chandra} X-ray data are from CCD3 (ACIS chip I3) using events in the 0.3-7 keV range. Tabulated quantities are: 
source name, J2000.0 X-ray position (R.A., Decl.), net counts and 
net counts error from {\em wavdetect} (accumulated in a 67,180 s exposure, rounded to the nearest integer,
background subtracted and PSF-corrected), 25\%, 50\% (median), and 75\%  photon quartile energies  
E$_{25}$, E$_{50}$, and E$_{75}$, 
K$_{s}$ magnitude of near-IR 2MASS counterpart, and 2MASS (2M) or radio (VLA) candidate counterpart identification
within a 2$''$ search radius. The offset (in arcsecs) between the X-ray 
and counterpart position is given in parentheses. A (v) following Net Counts error indicates that the source is likely variable 
as indicated by a variability 
probability P$_{var}$ $\geq$ 0.95 determined from the Kolmogorov-Smirnov (KS) statistic.
}
\tablenotetext{b}{The name IRS 3 corresponds to the infrared source identified by W93, which lies
                  $\approx$2.3$''$ northeast of the X-ray source.}
\tablenotetext{c}{The IRS 3 K magnitude is from W93.}
\tablenotetext{d}{The closest 2MASS source (2M J05470473+0021497)  lies at an offset of 2.34$''$ and K$_{s}$ $=$ 12.88.
                  A {\em Spitzer} IRAC source is visible and the 4.5 $\mu$m IRAC position  gives an  
                  offset of $\approx$0.42$''$ from  the  {\em Chandra} position.    }
\tablenotetext{e}{The X-ray source is offset by 0.27$''$ from radio source VLA J054705.367+002150.51. VLA coordinates are from
                  C. Carrasco-Gonz\'{a}lez 2009 (private communication).}
\tablenotetext{f}{Faint source not found by {\em wavdetect}. Net counts are measured within an extraction 
                     circle of radius 1$''$. Insufficient counts for variability analysis or calculation of
                     reliable E$_{25}$ and E$_{75}$ values.}
\tablenotetext{g}{A {\em Spitzer} IRAC source is visible and the  3.6 $\mu$m position gives an 
                  offset of $\approx$1.19$''$ from  the {\em Chandra} position.    }
\end{deluxetable}

\clearpage

\begin{deluxetable}{lllll}
\tabletypesize{\small}
\tablewidth{0pt}
\tablecaption{Spitzer IRAC Photometry: NGC 2071-IR\tablenotemark{a}}
\tablehead{
	\colhead{Object} &
	\colhead{[3.6]}	 &
	\colhead{[4.5]}  &
	\colhead{[5.8]}	 &
	\colhead{[8.0]} \\
	\colhead{     } &
	\colhead{(mag)} &
	\colhead{(mag)} &
	\colhead{(mag)} &
	\colhead{(mag)} 
	}
\startdata
IRS 1    & 8.359 $\pm$ 0.005& 6.246 $\pm$ 0.005& 4.576 $\pm$ 0.003& 2.819 $\pm$ 0.001 \\
IRS 2    & 7.747 $\pm$ 0.046& 6.382 $\pm$ 0.008& 5.581 $\pm$ 0.005& 4.717 $\pm$ 0.005 \\
IRS 3    & ...              & 7.250 $\pm$ 0.015& ...              & ...               \\
IRS 4    & 9.418 $\pm$ 0.035& 8.263 $\pm$ 0.007& 7.776 $\pm$ 0.005& 6.678 $\pm$ 0.007 \\
IRS 5\tablenotemark{b}  & ...              & ...              & ...              & ...               \\
IRS 6    & 8.609 $\pm$ 0.023& 8.052 $\pm$ 0.003& 7.619 $\pm$ 0.003& 6.966 $\pm$ 0.006 \\
IRS 7\tablenotemark{c}     & 8.546 $\pm$ 0.051& 6.705 $\pm$ 0.003& 5.776 $\pm$ 0.002& 4.654 $\pm$ 0.002 \\
IRS 8\tablenotemark{d}    & $>$11.35         & ... & ...           & ...               \\
\enddata

\tablenotetext{a}{
IRAC magnitudes were derived using  
a circular aperture of radius
r $=$ 2 pixels and a background annulus r $=$ 2 - 6 pixels 
(pixel size $\approx$1.22$''$). The
given magnitudes are background-subtracted and
aperture-corrected. Aperture correction factors are:~
1.213 (3.6 $\mu$m), 1.234 (4.5 $\mu$m),
1.379 (5.8 $\mu$m), 1.584 (8.0 $\mu$m).
Magnitude zero points (Z$_{0}$) for count rate units of DN s$^{-1}$ are:~
19.6642 (3.6 $\mu$m), 18.9276 (4.5 $\mu$m),
16.8468  (5.8 $\mu$m), 17.3909 (8.0 $\mu$m), where
[mag] = $-$2.5~log$_{10}$(DN s$^{-1}$) $+$ Z$_{0}$. 
Magnitude uncertainties are formal uncertainties 
(internal errors only).  The actual
uncertainties will in general be larger than the quoted
values due to systematic effects such as zero point 
uncertainties and other factors which can affect measurement
accuracy such as bright nearby nebulosity.
}
\tablenotetext{b}{No photometry measured. IRS 5 is 
                  located between the nearby bright  
                  sources IRS 1 and IRS 2 and is not clearly visible
                  in IRAC images.}
\tablenotetext{c}{The MIPS 24 $\mu$m data for  IRS 7 give a magnitude  
                  0.71 $\pm$ 0.03 mag.}
\tablenotetext{d}{Source appears nebulous and is 
                  extended in NE/SW direction. Possibly a close pair. 
                  Photometry is
                  uncertain and the lower limit is based on counts inside a r = 2 pixel
                  extraction circle with no background subtraction, 
                  centered at J054704.28$+$002137.9.}
\end{deluxetable}

\clearpage




\begin{deluxetable}{lllll}
\tabletypesize{\scriptsize}
\tablewidth{0pc}
\tablecaption{{\em Chandra} Spectral Fit Results 
   \label{tbl-1}}
\tablehead{
\colhead{Parameter}      &
\colhead{       }        &
\colhead{       }        &
\colhead{       } 
}
\startdata
Object                            & IRS 1               & IRS 1                         &  VLA 1      & IRS 6                \nl
Model                             & 2T APEC\tablenotemark{a}  & VPSHOCK\tablenotemark{a} &  1T APEC       & 1T APEC\tablenotemark{b}   \nl
N$_{\rm H}$ (10$^{22}$ cm$^{-2}$) & 3.62 [2.75 - 4.52] & 3.03 [2.20 - 4.52] & 8.68 [4.88 - 15.3] & 1.24 [0.37 - 2.31]  \nl
kT$_{1}$ (keV)                    & \{0.55\}\tablenotemark{c}               & 2.31 [1.31 - 5.99] & 2.22 [1.00 - 7.18] & 3.85 [1.47 - ... ]  \nl
norm$_{1}$ (10$^{-4}$)            & 2.60 [0.38 - 5.03] & 0.46 [0.23 - 0.82] & 1.18 [0.95 - 1.39] & 0.31 [0.17 - 0.65]  \nl
kT$_{2}$ (keV)                    & 3.31 [1.51 - 21.4] & ...                & ...                & ...                 \nl
norm$_{2}$ (10$^{-5}$)            & 2.82 [1.07 - 5.31] & ...                & ...                & ...                 \nl
E$_{line}$ (keV)                  & \{6.40\}           & \{6.40\}           & ...                & ...                 \nl
$\sigma_{line}$ (keV)             & \{0.05\}           & \{0.05\}           & ...                & ...                 \nl
norm$_{line}$ (10$^{-7}$)         & 6.89 [1.49 - 12.0] & 7.10 [2.43 - 12.5] & ...                & ...                 \nl 
Abundances                        & solar\tablenotemark{d}                  & solar\tablenotemark{e}  & solar    & solar               \nl
$\chi^2$/dof                      & 8.10/8             & 9.3/8              & 2.13/5             & 1.26/3              \nl
$\chi^2_{red}$                    & 1.01               & 1.16               & 0.43               & 0.42                \nl
F$_{\rm X}$ (10$^{-14}$ ergs cm$^{-2}$ s$^{-1}$)       & 2.99 (66.8)        & 2.84 (24.2)        & 2.54 (15.7)    & 2.65 (4.71)         \nl
F$_{\rm X,line}$ (10$^{-15}$ ergs cm$^{-2}$ s$^{-1}$)  & 7.01 (7.01)        & 8.04 (8.04)        & ...            &  ...                \nl
log L$_{\rm X}$ (ergs s$^{-1}$)   & 31.08              & 30.64              & 30.45              & 29.93               \nl
\enddata
\tablecomments{
Based on  fits of binned ACIS spectra using XSPEC v12.4.0. 
The spectra were rebinned to a minimum of
10  counts per bin. The tabulated parameters
are absorption column density (N$_{\rm H}$), plasma temperature (kT),
XSPEC normalization (norm), 
Gaussian line centroid energy (E$_{line}$), line width
($\sigma_{line}$ = FWHM/2.35), line normalization (norm$_{line}$).
For the APEC model, the XSPEC norm is related to the 
emission measure (EM) by EM = 4$\pi$10$^{14}$d$_{cm}^2$$\times$norm, 
where d$_{cm}$ is the stellar distance in cm. Quantities
enclosed in curly braces were held fixed during fitting.
Solar abundances are referenced to Anders \& Grevesse (1989).
Square brackets enclose 90\% confidence intervals and an ellipsis
means that the algorithm used to compute the confidence bound did not 
converge.
X-ray flux (F$_{\rm X}$) is the  observed (absorbed) value followed
in parentheses by the unabsorbed value in the 0.5 - 7.5 keV range.
The continuum-subtracted Gaussian line flux (F$_{\rm X,line}$) 
is measured in the 6.2 - 6.6 keV range. 
X-ray luminosity (L$_{\rm X}$) is the unabsorbed value in the
0.5 - 7.5 keV range. A distance of 390 pc is assumed.} 
\tablenotetext{a}{A fixed-width Gaussian line at 6.4 keV was included in the fit to model the faint
                  fluorescent Fe emission.}
\tablenotetext{b}{Due to low counts, the spectrum was binned to a minimum of 8 counts per bin.}
\tablenotetext{c}{The value of kT$_{1}$ is not tightly constrained. Values in the range 
                  kT$_{1}$ = 0.25 - 0.70 keV give nearly identical $\chi^2$ and lower
                  values of kT$_{1}$ converge to higher N$_{\rm H}$.}
\tablenotetext{d}{The solar abundance 2T APEC fit can be improved by allowing the Si abundance to vary. This yields
                  kT$_{1}$ = [0.55] keV (held fixed), N$_{\rm H}$ = 2.49 [1.44 - 3.73] $\times$ 10$^{22}$ cm$^{-2}$,
                  norm$_{1}$  = 3.04 [2.66 - 16.3] $\times$ 10$^{-5}$, 
                  kT$_{2}$ = 2.55 [1.53 - 7.06] keV
                  norm$_{2}$  = 3.42 [2.01 - 5.62] $\times$ 10$^{-5}$, 
                  Si = 5.1 [1.4 - 19.] $\times$ solar, 
                  $\chi^2$/dof = 4.94/7, and log L$_{\rm X}$ = 30.40 ergs s$^{-1}$. }
\tablenotetext{e}{The solar abundance VPSHOCK fit can be improved by allowing the Si abundance to vary. This yields
                  kT$_{1}$ = 3.27 [1.43 - 12.0] keV, N$_{\rm H}$ = 2.68 [1.20 - 3.66] $\times$ 10$^{22}$ cm$^{-2}$,
                  norm$_{1}$  = 2.92 [1.23 - 5.44] $\times$ 10$^{-5}$, Si = 2.2 [1.0 - 13.9] $\times$ solar, 
                  $\chi^2$/dof = 6.33/7, and log L$_{\rm X}$ = 30.44 ergs s$^{-1}$.}
\end{deluxetable}

\clearpage


\begin{figure}
\figurenum{1}
\includegraphics*[width=15.0cm,angle=0]{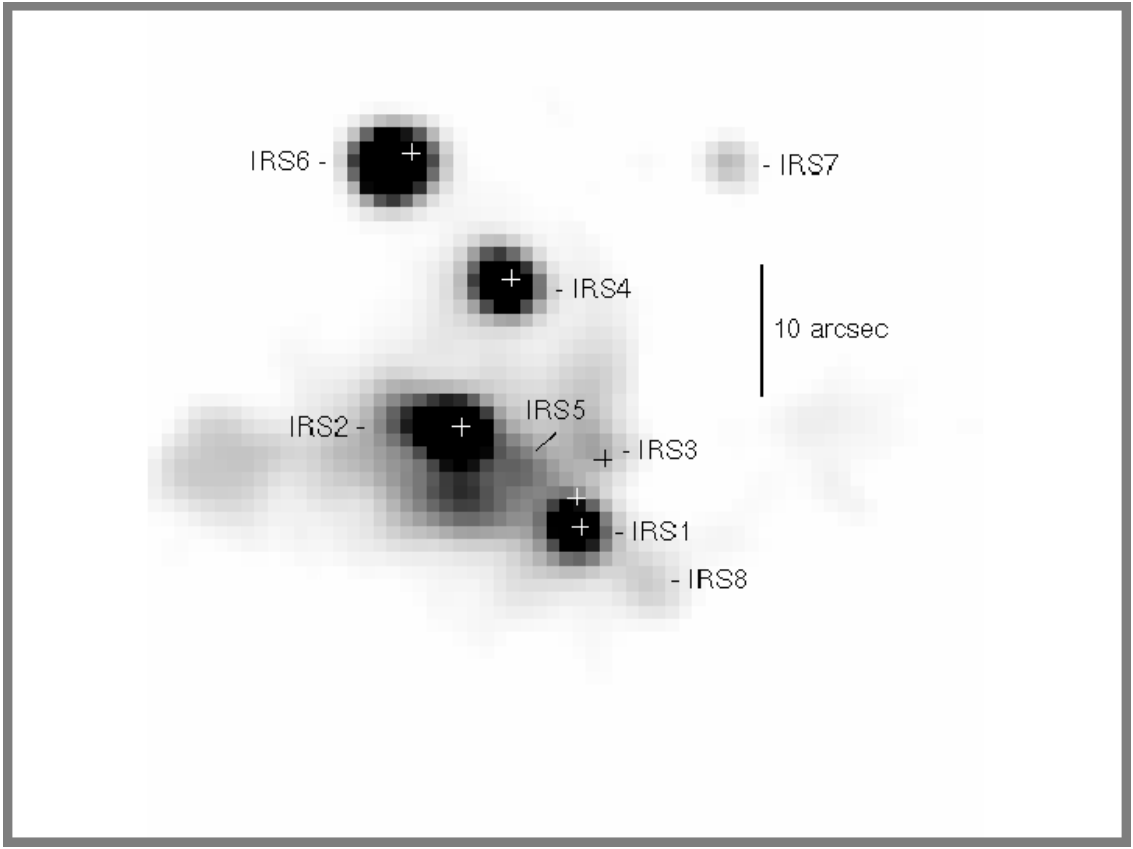}
\caption{A 2MASS K$_{s}$-band (2.16 $\mu$m) image of the region
         near IRS 1. Crosses mark positions of Chandra X-ray sources.
         Significant offsets between  Chandra and 2MASS
         positions exist for IRS 3 (2.$''$3) and
         IRS 6 (1.$''$7). See Table 1 for positions. No X-ray
         emission was detected by Chandra at the 2MASS positions of 
         IRS 5, IRS 7, and IRS 8.  
         Log intensity scale. North is up, east is left.}
\end{figure}

\clearpage

\begin{figure}
\figurenum{2}
\includegraphics*[width=15.0cm,angle=-90]{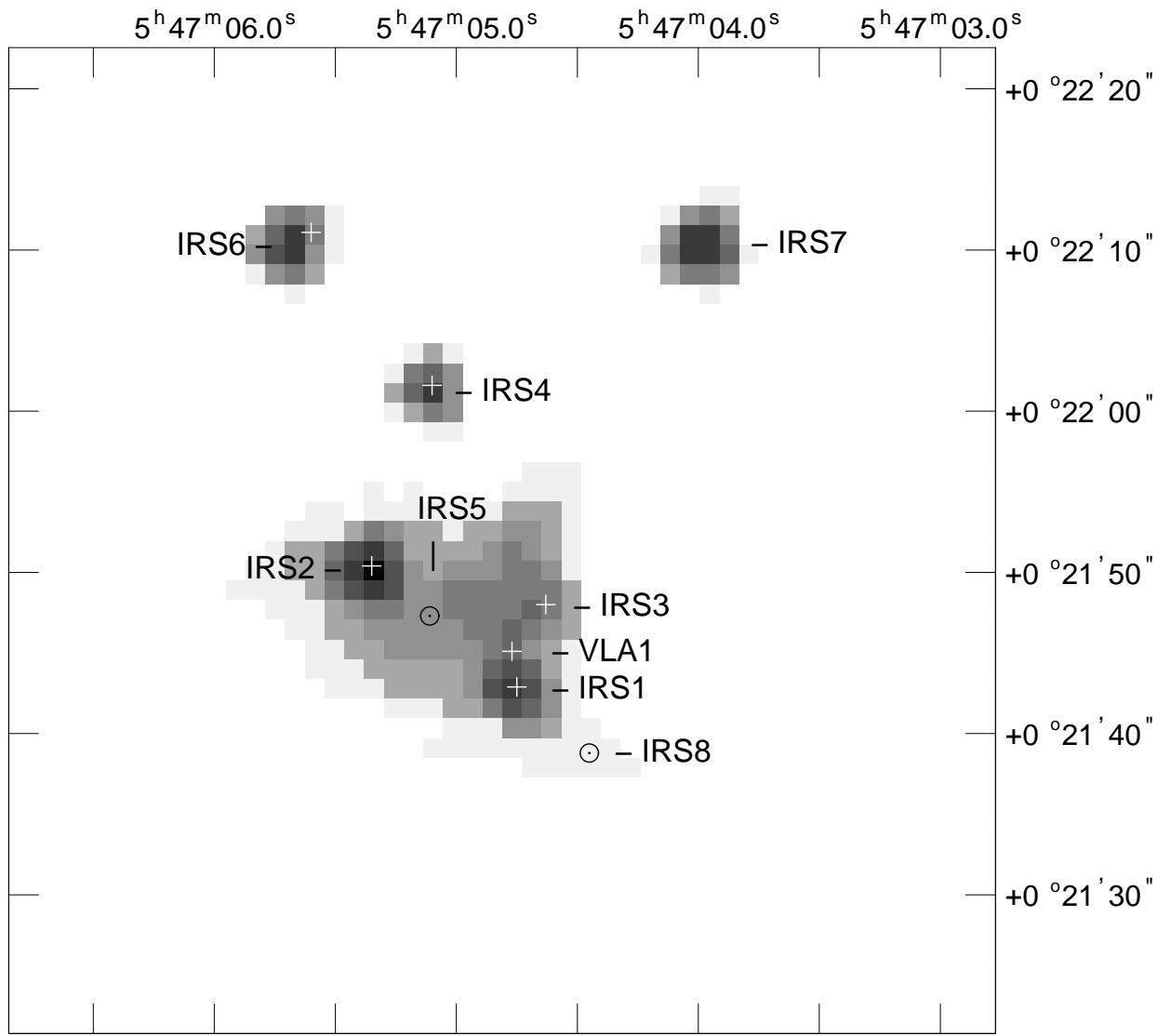}
\caption{An unsaturated  Spitzer  IRAC 3.6 $\mu$m (ch1) image of the region
         near IRS 1 (program 043). The image is a single 0.6 s frame. Crosses
         mark positions of Chandra X-ray sources. The positional offset
         between Chandra and IRAC for IRS 3 is only 0.$''$42, in better
         agreement than obtained with 2MASS (Fig. 1). 
         The positional offset between Chandra and IRAC for the binary
         system IRS 6  is  1.$''$2, again better than 
         obtained with 2MASS (Fig. 1).  IRS 5, IRS 7, 
         and IRS 8   were not detected by Chandra.  Circled dots   mark
         the 2MASS near-IR  positions of IRS 5 and IRS 8 for 
         clarity. Log intensity scale. Coordinates are J2000.}
\end{figure}

\clearpage

\begin{figure}
\figurenum{3}
\includegraphics*[width=11.0cm,angle=-90]{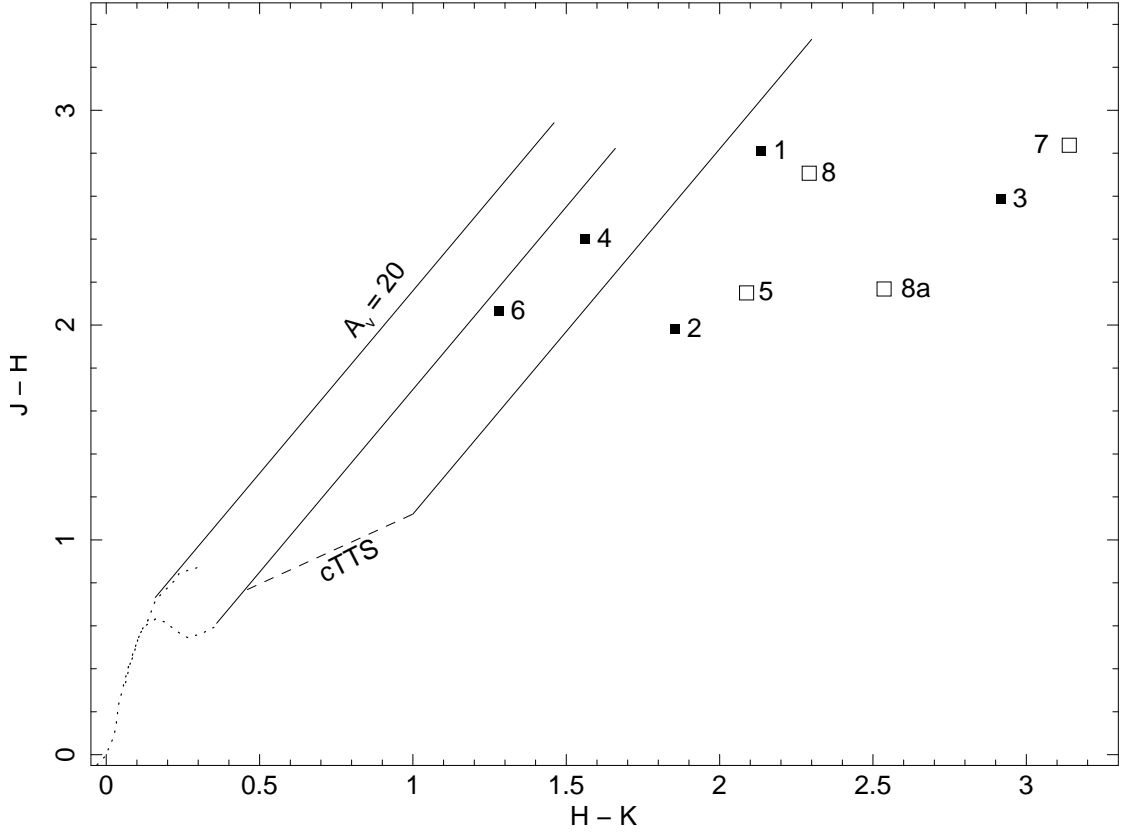}
\caption{Near-IR color-color diagram of NGC 2071-IR sources (Fig. 1).
         The colors are based on 
         the photometry of Tamura et al. (2007) transformed into
         the CIT  photometric system from the 2MASS system
         using the transformation of Carpenter (2001). Solid squares
         are sources detected by {\em Chandra} (Table 1) and open squares
         are non-detections. The dotted lines at the
         lower left are  the unreddened colors of dwarf and giant
         stars based on the intrinsic colors of  Bessell \& Brett (1988) 
         converted  into the CIT system. The dashed line is 
         the classical T Tauri star locus from Meyer et al. (1997)
         representing the range of intrinsic IR excess for young
         stars with evidence for accreting circumstellar disks in Taurus.
         The three solid lines show representative  reddening vectors  
         based on Cohen et al. (1981)  for A$_{\rm V}$ = 20 mag as measured in
         the CIT system. 
         The colors of IRS 4 and IRS 6 are consistent
         with reddened classical T Tauri stars while the rest of the sources 
         appear to have more complex circumstellar environments.}
\end{figure}
\clearpage

\begin{figure}
\figurenum{4}
\includegraphics*[width=11.0cm,angle=-90]{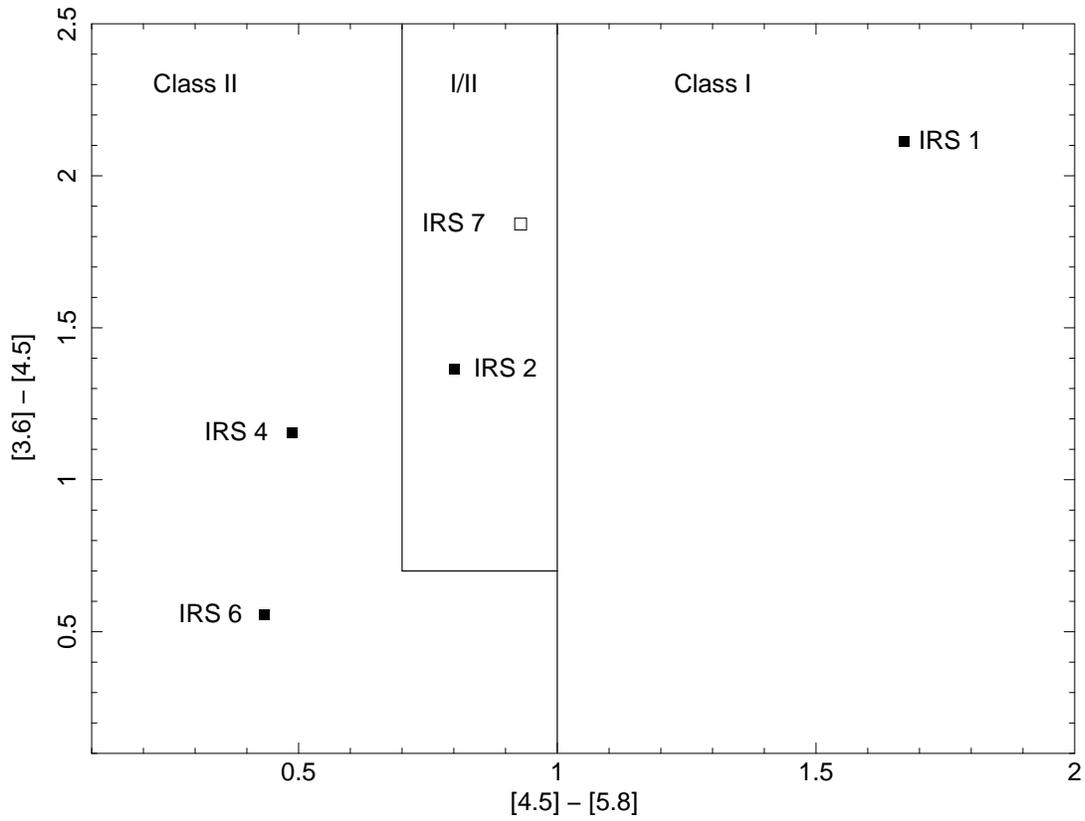}
\caption{IRAC color-color diagram for sources in NGC 2071-IR
         with measured photometry in all four IRAC bands (Table 2).
         Solid squares denote sources detected by {\em Chandra}
         and the open square marks the undetected object IRS 7.
         The  solid lines show the boundaries between class I,
         ~class I or possibly heavily-reddened class II, and 
         class II YSOs   based on the criteria 
         of Gutermuth et al. (2008).}
\end{figure}
\clearpage

\begin{figure}
\figurenum{5}
\includegraphics*[width=13.0cm,angle=-90]{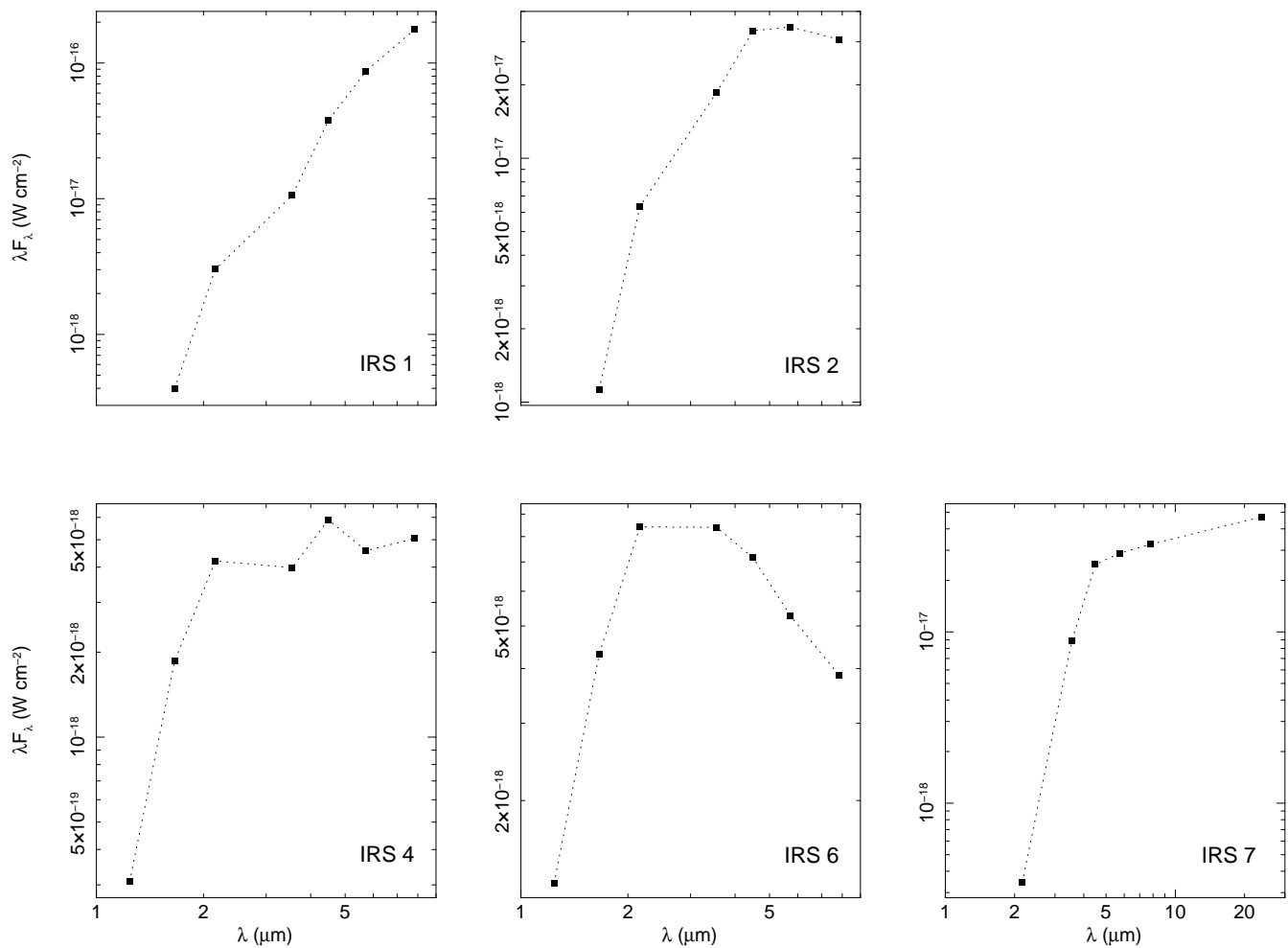}
\caption{Infrared spectral energy distributions of sources in 
         NGC 2071-IR with reliable near-IR and {\em Spitzer}
         data. The near-IR data are from {\em 2MASS}. {\em Spitzer}
         IRAC  data are plotted at 3.55, 4.49, 5.72, and 7.83 $\mu$m.
         A MIPS-24 measurement is shown for IRS 7 at 23.84 $\mu$m. }
\end{figure}
\clearpage

\begin{figure}
\figurenum{6}
\includegraphics*[width=15.0cm,angle=-90]{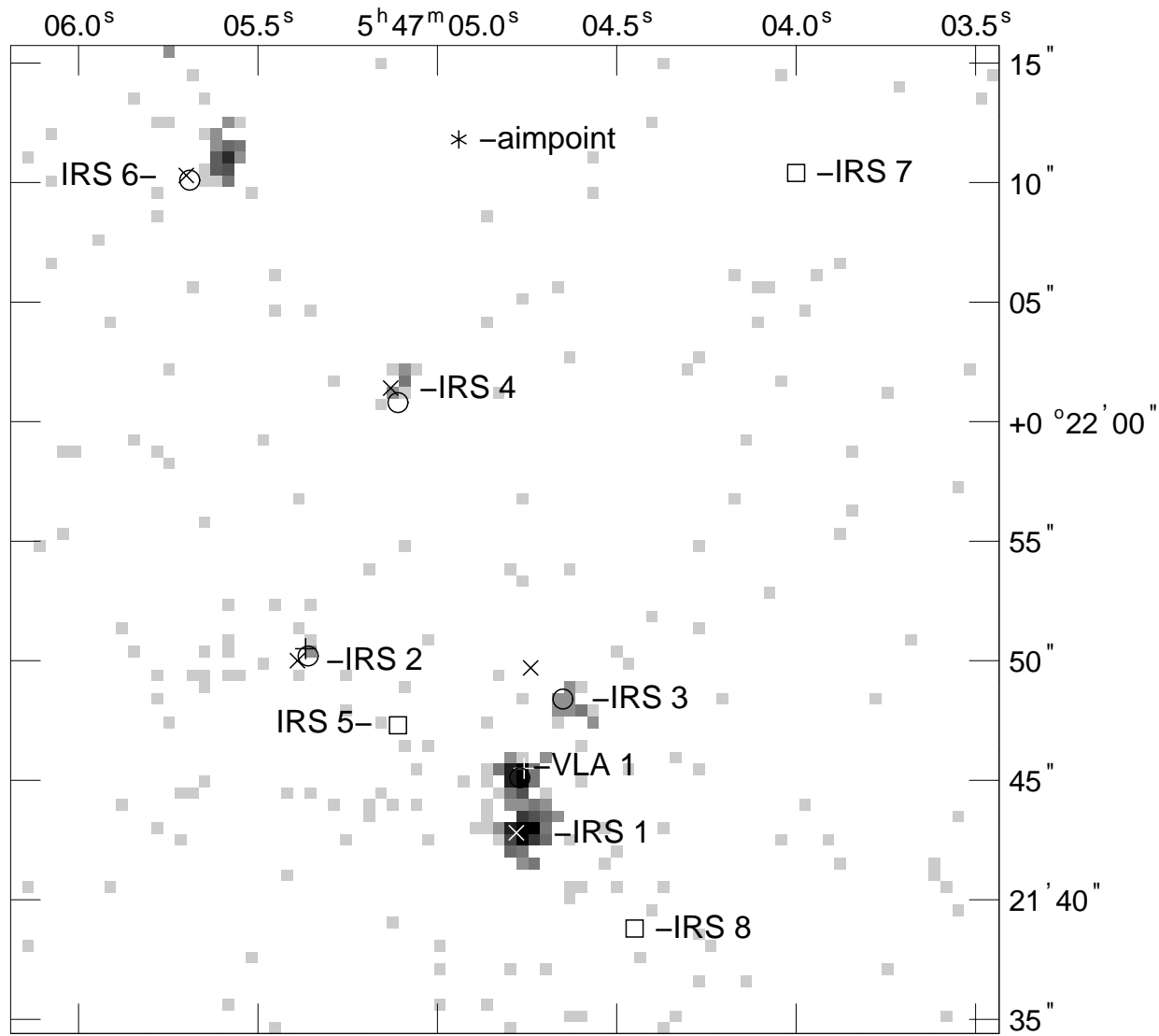}
\caption{{\em Chandra} ACIS-I image of the central region of the 
infrared cluster  NGC 2071-IR (0.3 - 7 keV, log intensity scale, 
0.$''$49 pixels, J2000. coordinates). An asterisk marks the 
{\em Chandra} aimpoint. The crosses ($\times$) mark 
2MASS source positions of near-IR sources with {\em Chandra} 
counterparts (IRS 1,2,3,4, and 6).
and open squares  mark 2MASS positions of the undetected cluster core 
sources IRS 5,7, and 8.
Circles show the near-IR positions of sources IRS 2, 3, 4, and 6
obtained by applying the offsets given in Walther et al. (1993)
to the 2MASS position of IRS 1. Plus signs ($+$) mark the 
{\em VLA} radio positions of VLA 1 and IRS 2A (C. Carrasco-Gonzalez
2009, priv. communication).}
\end{figure}

\clearpage

\begin{figure}
\figurenum{7}
\includegraphics*[width=15.0cm,angle=-90]{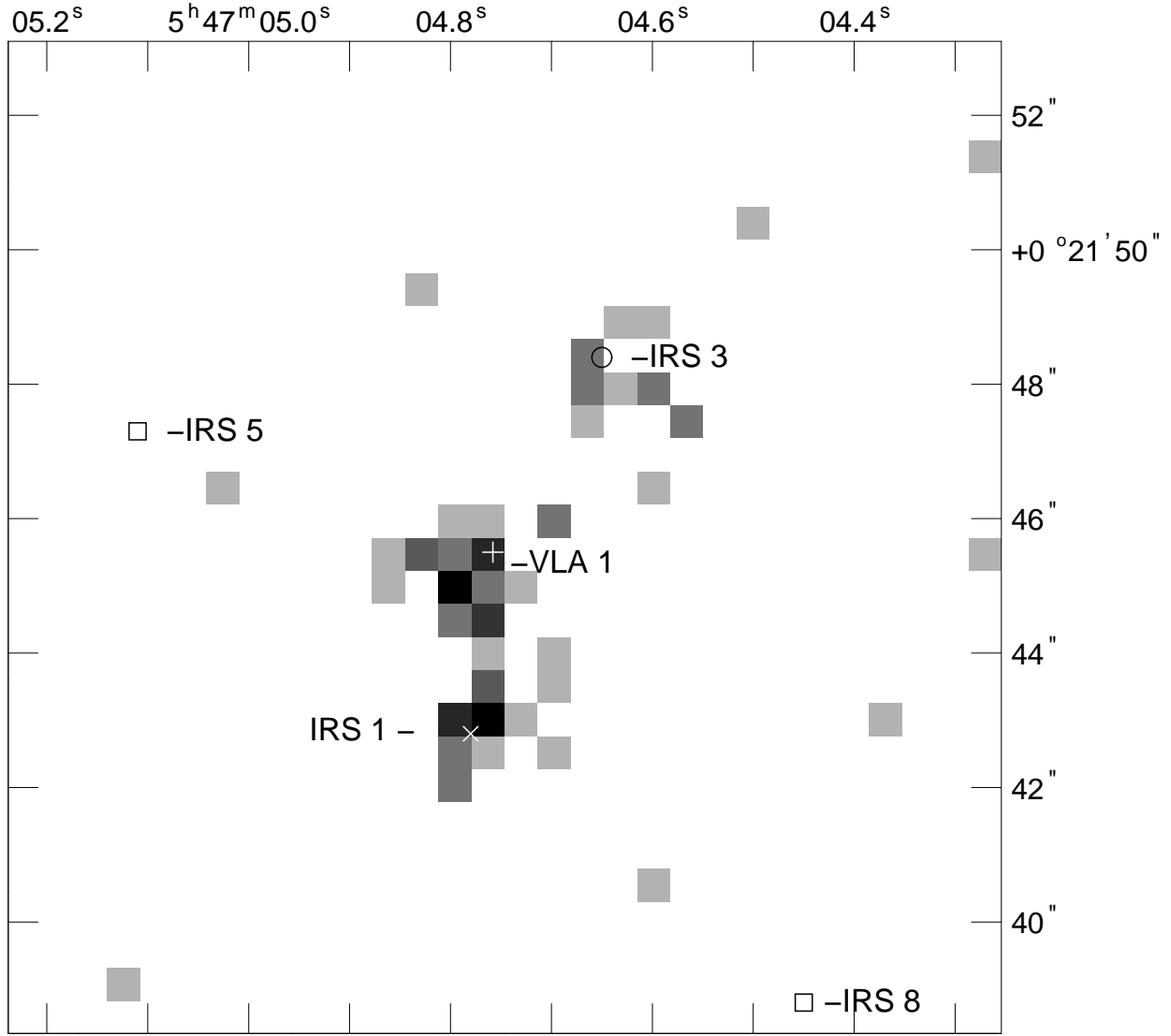}
\caption{Zoomed view of the {\em Chandra} ACIS-I image 
 near  IRS 1 in the hard energy range
 (4 - 7 keV, log intensity scale, 0.$''$49 pixels,
 J2000. coordinates). Symbols are the same as in Fig. 6.
 Hard emission is detected from IRS 1, IRS 3, and VLA 1.
 The near-IR sources IRS 5 and IRS 8 were undetected by {\em Chandra}.}
\end{figure}

\clearpage

\begin{figure}
\figurenum{8}
\includegraphics*[width=11.5cm,angle=-90]{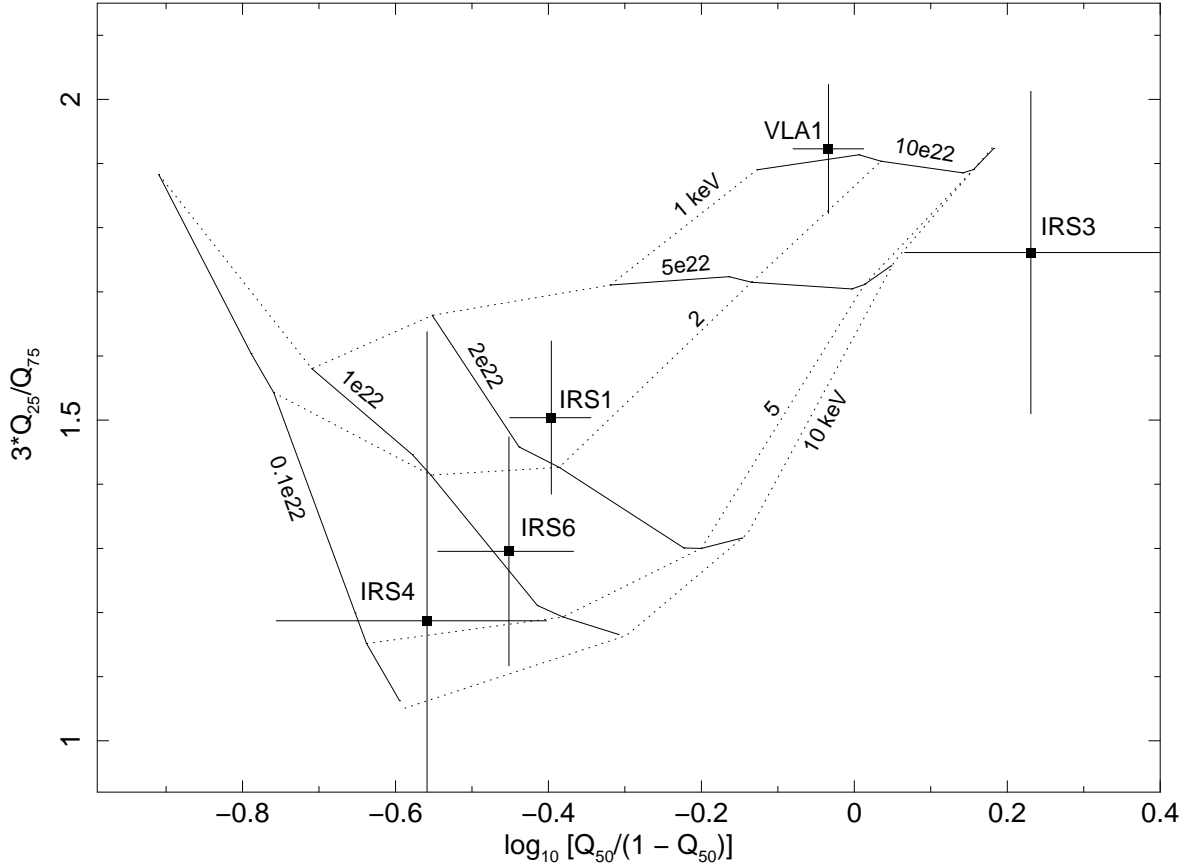}
\caption{Quantile plot of the X-ray sources in NGC 2071-IR.
 The source IRS 2 (4 net counts) is too faint for 
 quantile  analysis and is  not shown. The quantile values 
 are defined using the relation 
 Q$_{x}$ = (E$_{x\%}$ - E$_{lo}$)/(E$_{up}$ - E$_{lo}$)
 (Hong et al. 2004), where E$_{lo}$ = 0.3 keV and
 E$_{up}$ = 7.0 keV in this study. Solid lines show
 loci of constant absorption column density 
 N$_{\rm H}$ = (0.1, 1, 2, 5, 10) $\times$ 10$^{22}$ cm$^{-2}$
 and dotted lines show loci of constant plasma energy 
 kT =  1, 2, 5, and 10 keV. The grid of 
 N$_{\rm H}$ and kT values was generated using
 simulations based on a solar-abundance 1T APEC thermal
 plasma model and the 
 response and auxiliary response files for IRS 1.}
\end{figure}

\clearpage

\begin{figure}
\figurenum{9}
\epsscale{1.0}
\plottwo{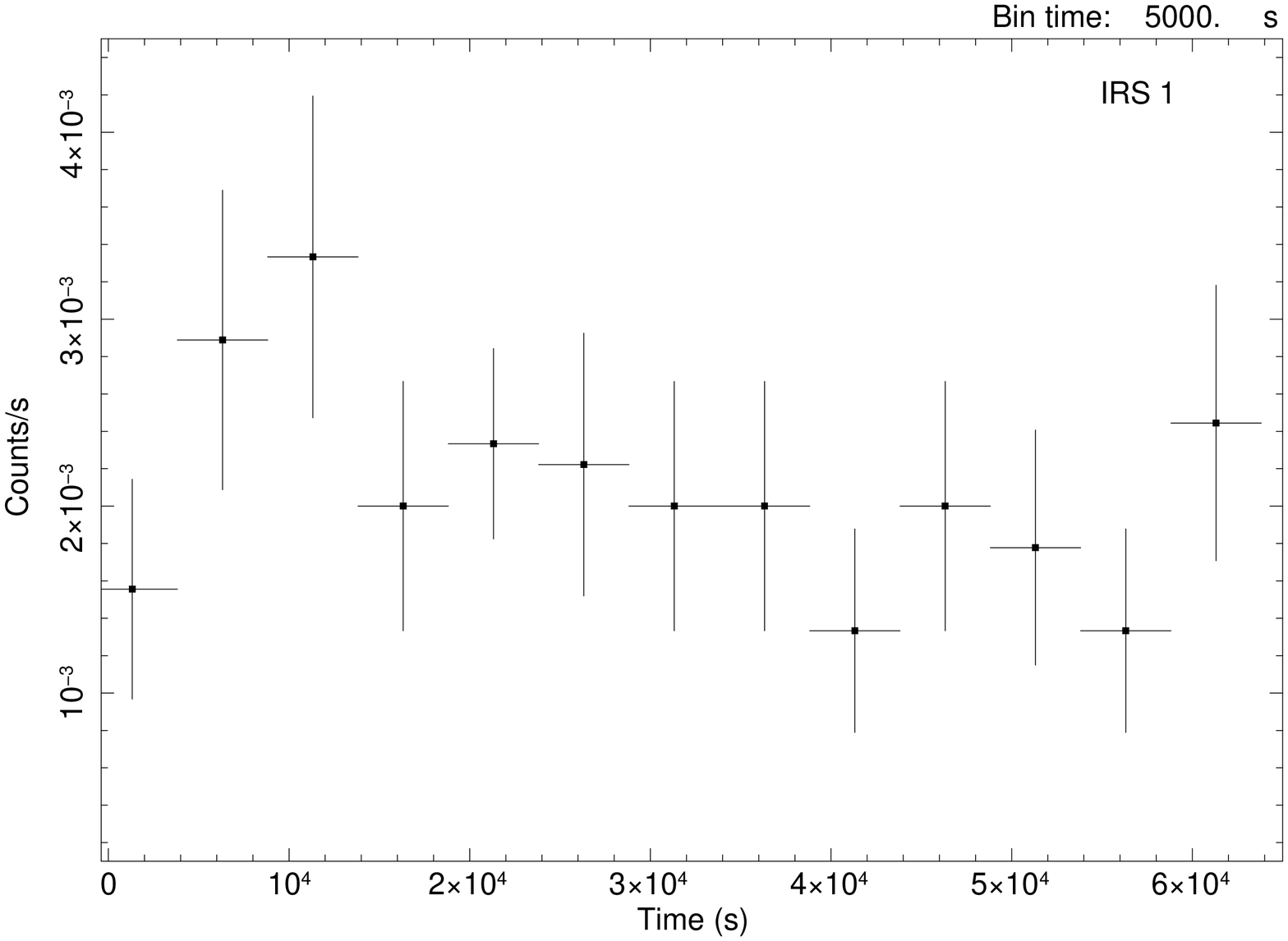}{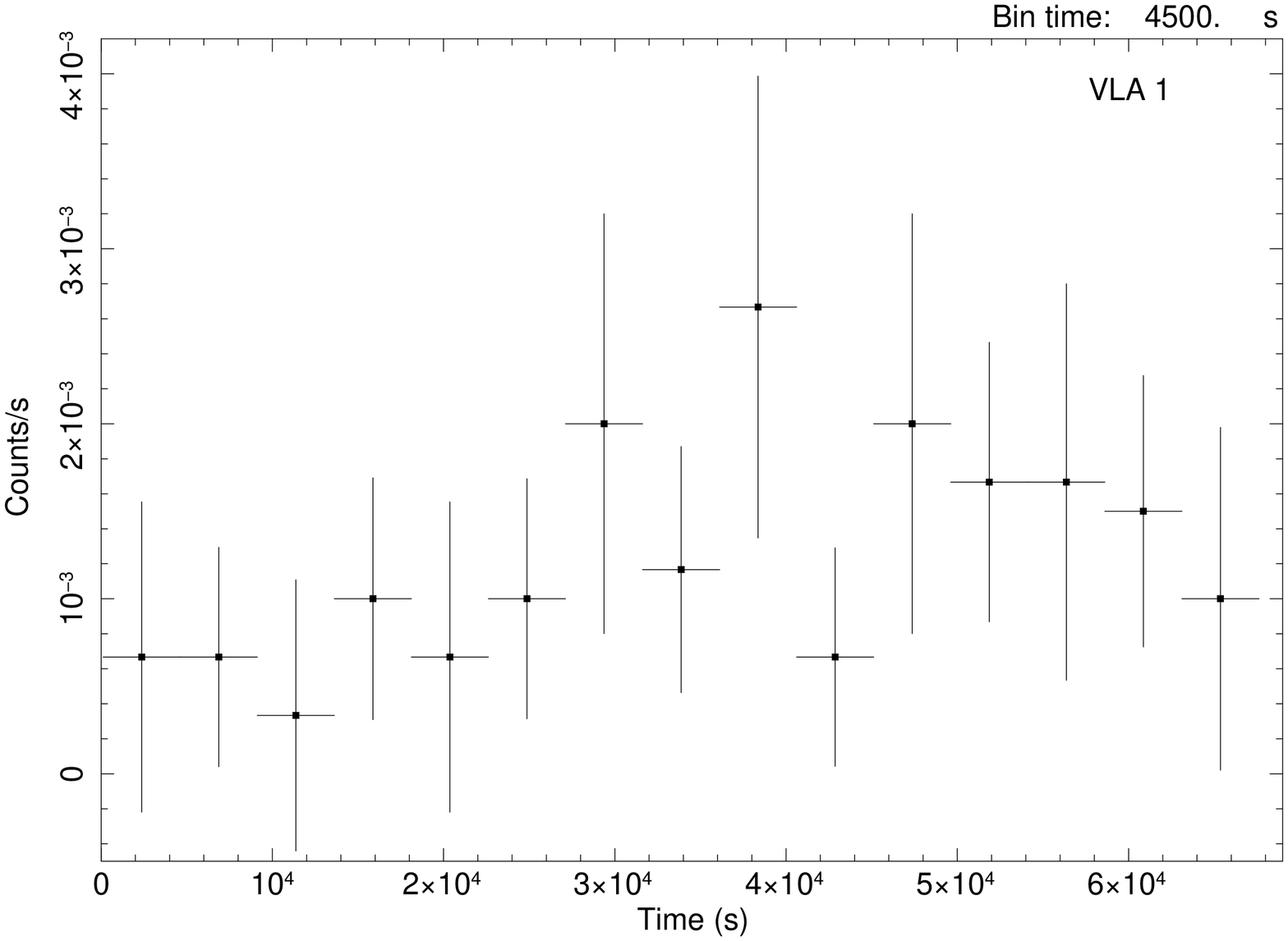}
\caption{
{\em Left}:~{\em Chandra} broad-band (0.3 - 7 keV) ACIS-I 
lightcurve of IRS 1 binned at 5000 s intervals.
The mean count rate is 2.02 $\pm$ 0.69 ($\pm$1$\sigma$) c ksec$^{-1}$.
The KS test gives a probability of constant count rate
P$_{\rm const}$ = 0.11, so variability is not demonstrated
with high confidence.
 ~{\em Right}:~{\em Chandra} broad-band (0.3 - 7 keV) ACIS-I 
lightcurve of VLA 1  binned at 4500 s intervals.
The mean count rate is 1.33 $\pm$ 0.75 ($\pm$1$\sigma$) c ksec$^{-1}$.
The KS test  gives 
P$_{\rm const}$ = 0.02, so variability is likely.
} 
\end{figure}
\clearpage

\begin{figure}
\figurenum{10}
\includegraphics*[width=12.0cm,angle=-90]{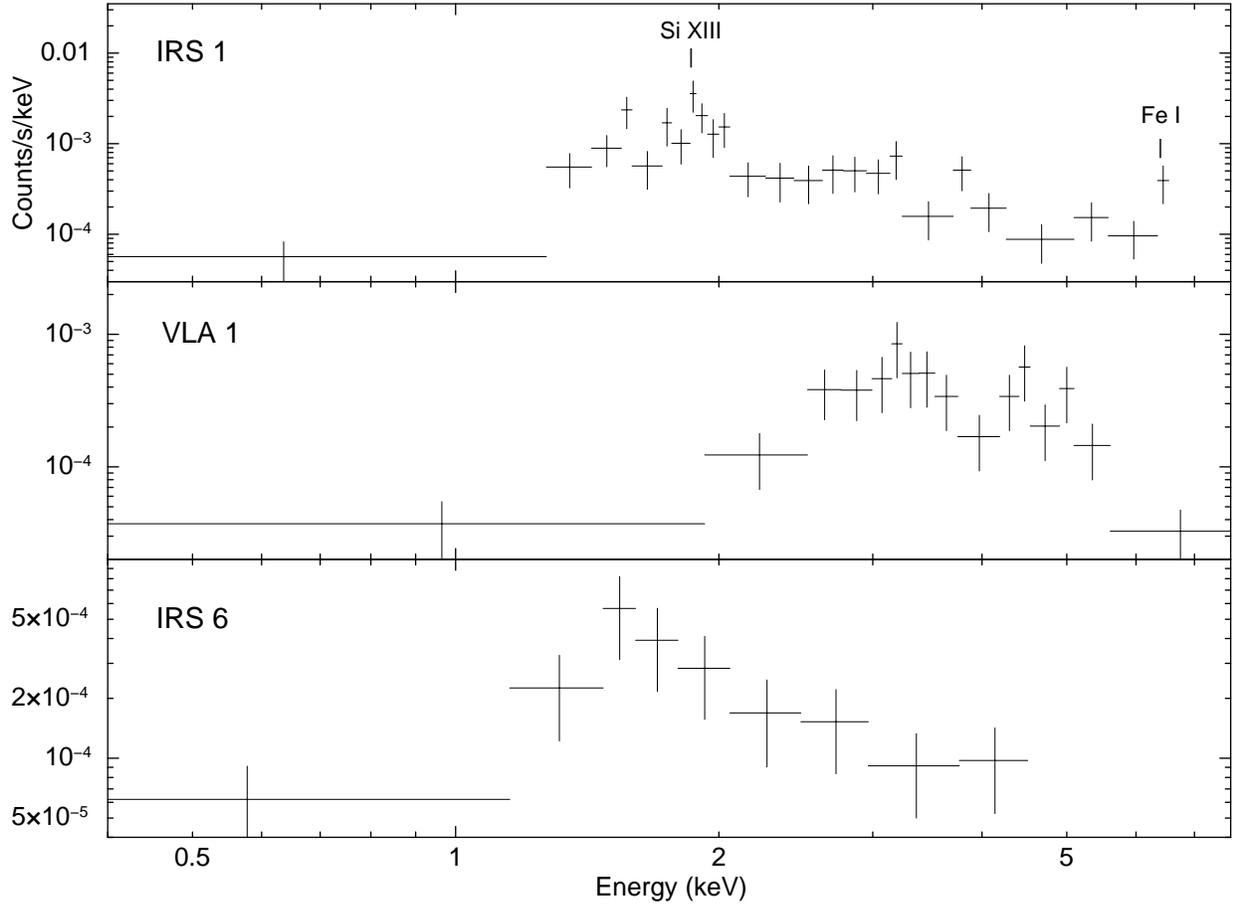}
\caption{ACIS-I spectra  of IRS 1, VLA 1, and IRS 6 binned to a  minimum of
5 counts per bin. } 
\end{figure}

\clearpage

\begin{figure}
\figurenum{11}
\includegraphics*[width=9.5cm,angle=-90]{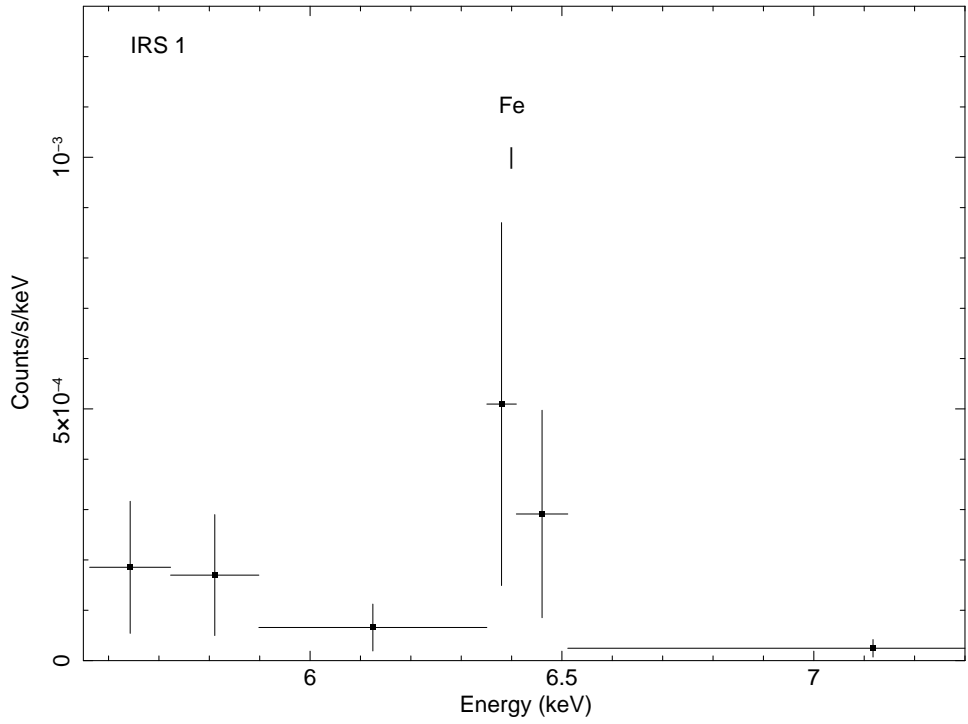}
\caption{ACIS-I spectrum  of IRS 1 binned to a  minimum of
2 counts per bin, showing the region near the fluorescent 
iron line.  Solid squares are data points.}
\end{figure}

\clearpage

\begin{figure}
\figurenum{12}
\includegraphics*[width=9.5cm,angle=-90]{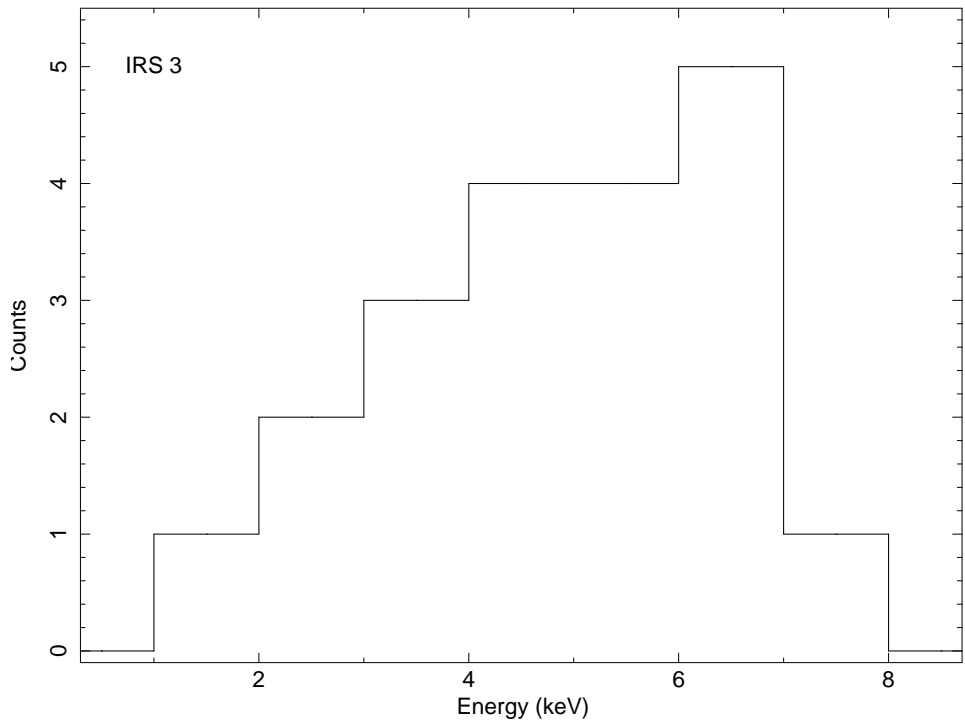}
\caption{ACIS-I event energy histogram of the hard X-ray source
         IRS 3 using events in the 0.3 - 7 keV range, binned at 
         1 keV intervals. The extraction region contained 20 events,
         of which 2 events are likely background. The median event
         energy is E$_{50}$ = 4.52 keV.}
\end{figure}

\clearpage

\end{document}